\documentclass[prb,prd,showpacs,preprintnumbers,amsmath,amssymb,eqsecnum,nofootinbib]{revtex4}

\usepackage{graphicx}
\usepackage{dcolumn}
\usepackage{bm}
\usepackage{color}

\newcommand{\be}{\begin{equation}}
\newcommand{\ee}{\end{equation}}
\newcommand{\bea}{\begin{eqnarray}}
\newcommand{\eea}{\end{eqnarray}}

\newcommand{\bi}{\begin{itemize}}
\newcommand{\ei}{\end{itemize}}
\newcommand{\benu}{\begin{enumerate}}
\newcommand{\eenu}{\end{enumerate}}
\newcommand{\nn}{\nonumber}
\newcommand{\noin}{\noindent}

\def \psl {p \kern-.45em{/}}
\def \qsl {q \kern-.45em{/}}
\def \ksl {k \kern-.55em{/}}
\def \lsl {l \kern-.40em{/}}
\def \ssl {s \kern-.45em{/}}
\def \asl {a \kern-.45em{/}}
\def \bsl {b \kern-.45em{/}}

\begin{document}

\preprint{MISC-2014-03}
\preprint{YNU-HEPTh-14-102} 

\title{Higgs boson production in $e$ and real $\gamma$ collisions
 }

\author{Norihisa WATANABE}
\email{norihisa@post.kek.jp}
\author{Yoshimasa KURIHARA}%
 \email{yoshimasa.kurihara@kek.jp}
\affiliation{
High Energy Accelerator Research Organization (KEK)\\  
Tsukuba, Ibaraki 305-0801, JAPAN
}%

\author{Tsuneo UEMATSU}
\email{uematsu@scphys.kyoto-u.ac.jp}
\affiliation{Institute for Liberal Arts and Sciences, Kyoto University, Kyoto 606-8501, Japan\\
        and Maskawa Institute, Kyoto Sangyo University, Kyoto 603-8555, Japan 
}%

\author{Ken SASAKI}
\email{sasaki@ynu.ac.jp}
\affiliation{Dept. of Physics, Faculty of Engineering,  Yokohama National University,  
 Yokohama 240-8501, JAPAN 
\vspace{2cm}}%

\date{July 24, 2014}

\begin{abstract}
We  investigate the  Standard Model Higgs boson production in  $e^-\gamma$ collisions. The electroweak one-loop 
contributions to the scattering amplitude for $e^-\gamma\rightarrow e^-H$ are calculated and  
expressed in analytical form. We analyze  the  cross section for the Higgs boson production in $e^-\gamma$ collisions  for each 
combination of polarizations of the initial electron and photon beams. The feasibility of observing the Higgs boson 
in $e^-+\gamma\rightarrow e^-+b+{\overline b}$ channel is examined. 

\end{abstract}

\maketitle

\section{Introduction}

A Higgs boson with mass about 125 GeV was discovered by ATLAS and CMS at LHC~\cite{HiggsLHC} and 
its spin, parity, and its couplings to other particles have been examined~\cite{SpinParity}. For further detailed studies of its properties, a new accelerator facility, a linear $e^+e^-$ collider, which offers much cleaner experimental collisions, is attracting growing attention~\cite{ILC}. Along with $e^+e^-$ collider, other options such as $e^-e^-$, $e^-\gamma$ and 
$\gamma\gamma$ colliders have also been discussed. See Refs.~\cite{DeRoeck}-\cite{GK} and the references therein. 
Each option for colliders will provide interesting topics to study, such as the detailed measurement of the Higgs boson properties and the quest for the new physics beyond the Standard Model (SM). An $e^-e^-$ collider is easier 
to build than an $e^+e^-$ collider and may stand as a potential candidate before  positron sources with 
high intensity are available. The $e^-\gamma$ and $\gamma\gamma$ options are based on  
$e^-e^-$ collisions, where one or two of the electron beams are converted to the photon beams.

In this paper we investigate the production of  the SM Higgs boson ($H$)  in an  $e^-\gamma$ collider\footnote{A part of this work has been reported elsewhere~\cite{KWSUPL}.}. We examine the reaction $e^-\gamma\rightarrow e^-H$ at the one-loop level in the electroweak interaction. Particularly, we are interested in the contribution from the two-photon fusion process $\gamma^*\gamma\rightarrow H$ which is described by the ``so-called" transition form factor of the Higgs boson~\cite{KWSUPL}. 
One of the advantages of linear colliders is that large polarization can be obtained for both beams.
We 
analyze the Higgs boson production cross section in $e^-\gamma$ collisions for each 
combination of polarizations of the initial electron and photon beams and discuss 
the feasibility of observing the Higgs bosons.

In fact, the Higgs boson production in $e^-\gamma$ collisions was investigated by Gabrielli, Ilyin and Mele some time ago 
before the Higgs boson was discovered~\cite{Gabrielli} (see also Ref.\cite{Resolved}). 
They surveyed the reaction $e^-\gamma\rightarrow e^-H$ in the   center-of-mass energy range
 $\sqrt{s}=(0.5-2)$ TeV and $m_h=(80-700)$ GeV.   Comparing their results 
 (at $\sqrt{s}=500$ GeV and $m_h=120$ GeV) with ours (at $\sqrt{s}=500$ GeV and $m_h=125$ GeV), 
 we found  some differences between the two.  
At the one-loop level in the electroweak interaction,  four groups of Feynman diagrams,   ``$\gamma^*\gamma$" ,  ``$Z^*\gamma$", ``$W\nu_e$" and  ``$Ze$" (which are defined in Sec.II), contribute to the reaction $e^-\gamma\rightarrow e^-H$ .  Although their result on the ``$\gamma^*\gamma$" contribution 
 is consistent with ours,  the contributions from ``$Z^*\gamma$" and ``$W\nu_e$"  were predicted to be much less than  ours. Indeed, it was reported in Ref.~\cite{Gabrielli}  that  the ``$\gamma^*\gamma$" contribution was dominant in the total cross section and thus the interference effect among different groups of diagrams was rather small. We find that the ``$Z^*\gamma$" contribution becomes  approximately  of the same magnitude as the one from ``$\gamma^*\gamma$" at 
 $\sqrt{s}$=500 GeV. Also,  for the case when the initial electron beam is left-handed, the ``$W\nu_e$" contribution prevails over the  ``$\gamma^*\gamma$" at $\sqrt{s}=500$ GeV.  We will show that  the interferences between  ``$\gamma^*\gamma$" and ``$Z^*\gamma$" and between ``$\gamma^*\gamma$" and  ``$W\nu_e$", which work destructively or constructively depending on the polarizations of the initial beams, are important factors affecting the behaviors of both the differential cross section and the cross section of the Higgs production. To make differences clear, we give explicit expressions for the results of our calculations. 

In the next section, we classify the one-loop diagrams for the reaction $e^-\gamma\rightarrow e^-H$  into four groups. The contribution to the scattering amplitude from each group of the diagrams is evaluated in unitary gauge and expressed in analytical form. 
In Sec. \ref{Section3},  the dependence of  the reaction  on the 
polarizations of the initial electron and photon beams is emphasized. 
Both the differential cross section and the cross section for $e^-\gamma\rightarrow e^-H$ are examined 
in each case of  polarizations of the initial beams. 
In Sec.\ref{Section4} we consider the case when a high-intensity photon beam is produced by laser light 
backward scattering off a high-energy electron beam and we  analyze
the Higgs boson production in $e^-\gamma$ collisions using an $e^-e^-$ collider machine. 
The final section is devoted to the conclusions. For completeness we add three Appendices where the 
Feynman rules we use are enumerated in Appendix \ref{FeynmanRules}, the analytical expressions of the relevant scalar one-loop integrals are given in Appendix \ref{OLI} and the contributions from the interference terms are written down 
in Appendix \ref{Interference}.

\section{Higgs boson production in  $e^-\gamma$ collisions \label{Section2}}
We analyze the Higgs production in  a $e^-$ and real $\gamma$ collision experiment, 
\bea
  e^-(k_1) +\gamma(k_2) \rightarrow 
 e^-(k_1') +H(p_h)~. \label{HiggsProduction}
\eea
The Higgs boson we consider is the one in the SM. The relevant Feynman diagrams for this process start not at  tree-level but at  the one-loop level in the electroweak interaction. We calculate the relevant one-loop diagrams in unitary gauge using dimensional regularization  which respects electromagnetic gauge invariance. In the unitary gauge, only the physical particles appear and ghosts and Goldstone bosons are absent. The  gauge boson propagators in unitary gauge  and  the relevant Feynman rules for the three- and four-point vertices which we use for this work are summarized in Appendix \ref{FeynmanRules}. The one-loop diagrams which contribute to the reaction (\ref{HiggsProduction}) are classified into four groups:  $\gamma^*\gamma$ fusion diagrams (Fig.\ref{ggfusiontquark},~\ref{ggfusionWboson}),  $Z^*\gamma$ fusion diagrams, ``$W\nu_e$" diagrams (Fig.\ref{Wrelated}) and  ``$Ze$" diagrams (Fig.\ref{Zrelated}). 

Since $k_2$ is the momentum of a real photon,  we have $k_2^2=0$ and $k_2^\beta \epsilon_\beta(k_2)=0$, where $\epsilon_\beta(k_2)$ is the photon polarization vector.  We set $q=k_1-k'_1$. Assuming that  electrons are massless so that $k_1^2={k'_1}^2=0$, we introduce the following Mandelstam variables:
\bea
s&=&(k_1+k_2)^2=2k_1\cdot k_2~,\quad t=(k_1-k_1')^2=q^2=-2k_1\cdot k_1',\\
u&=&(k_1-p_h)^2=-2k_1'\cdot k_2=m_h^2-s-t~.
\eea
where $p_h^2=m_h^2$ with $m_h$ being the Higgs boson mass.

\subsection{Virtual photon--real photon fusion diagrams}

\begin{figure}[htbp]
 \begin{tabular}{c}
 \begin{minipage}{0.5\hsize}
  \begin{center}
   \includegraphics[width=50mm]{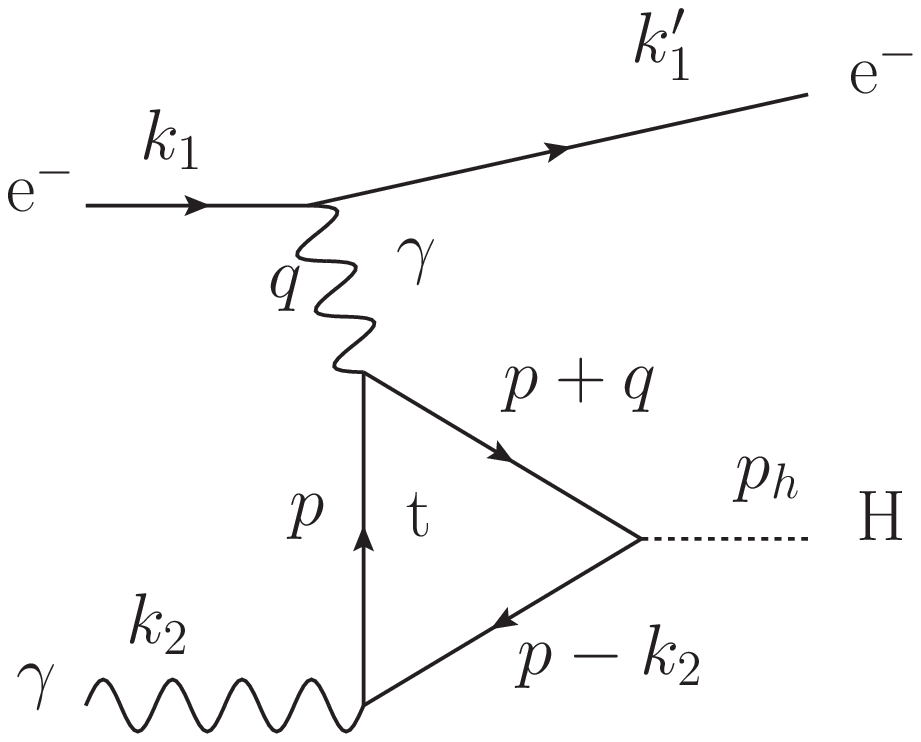}
  \end{center}
 \end{minipage}
 \begin{minipage}{0.5\hsize}
  \begin{center}
   \includegraphics[width=50mm]{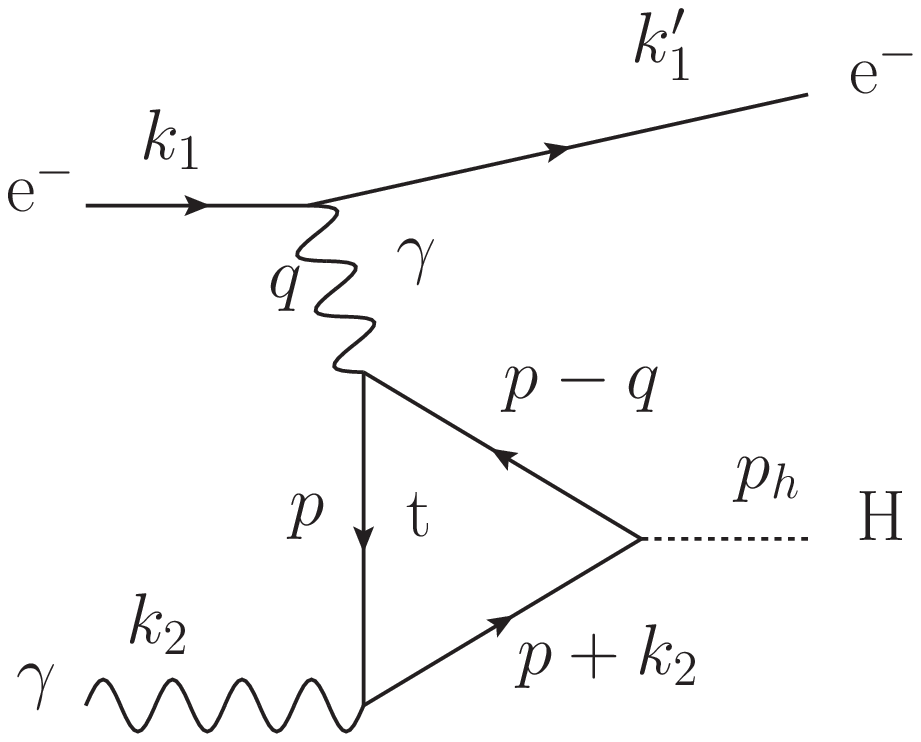}
  \end{center}
 \end{minipage}
 \end{tabular}
\caption{ $\gamma^*\gamma$ fusion diagrams: top-quark loop contributions}
\label{ggfusiontquark}
\end{figure}

\begin{figure}[htbp]
 \begin{tabular}{c}
 \begin{minipage}{0.33\hsize}
  \begin{center}
   \includegraphics[width=50mm]{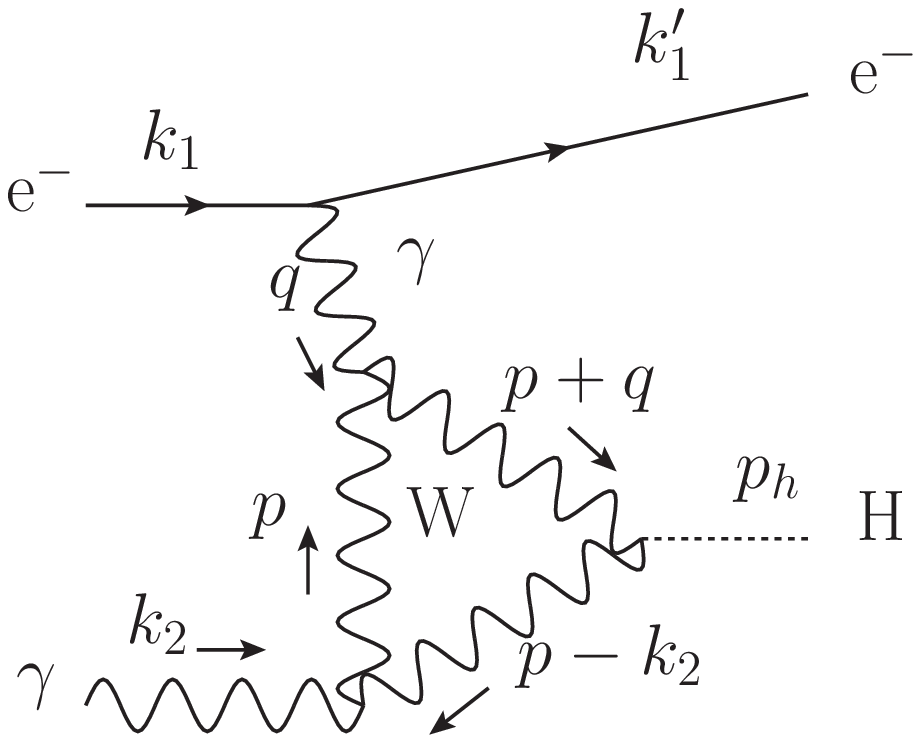}
  \end{center}
 \end{minipage}
 \begin{minipage}{0.33\hsize}
 \begin{center}
  \includegraphics[width=50mm]{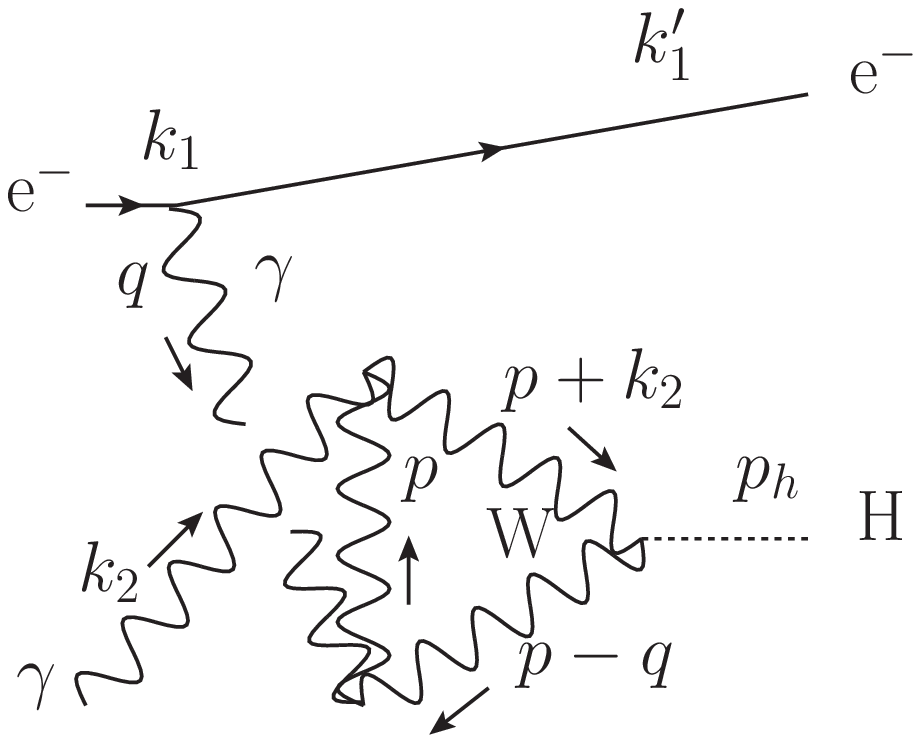}
 \end{center}
 \end{minipage}
 \begin{minipage}{0.33\hsize}
 \begin{center}
  \includegraphics[width=50mm]{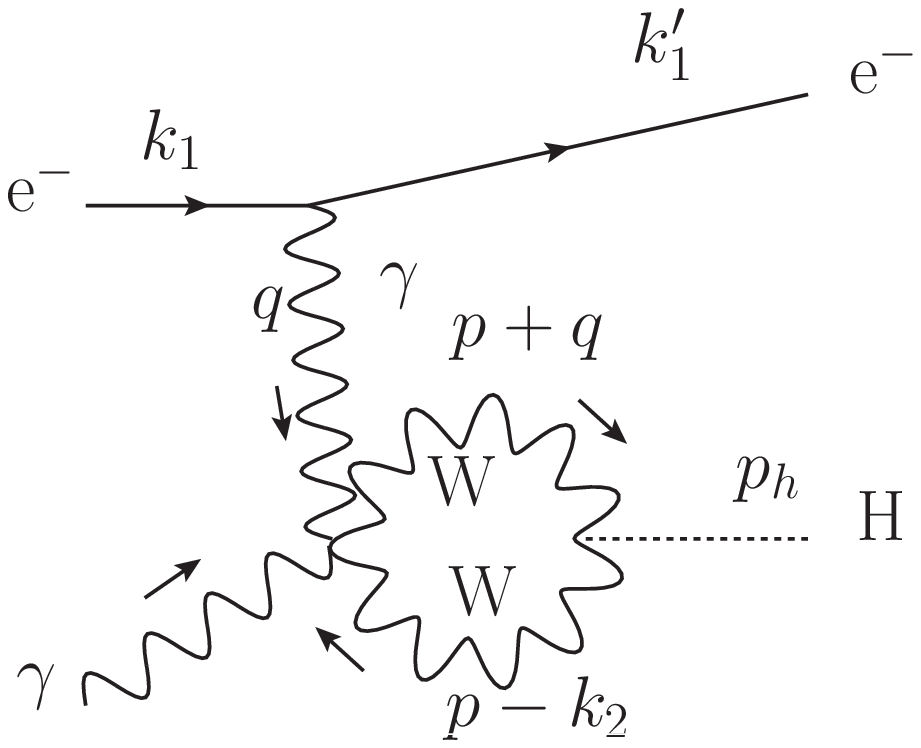}
 \end{center}
 \end{minipage}
 \end{tabular}
 \caption{  $\gamma^*\gamma$ fusion diagrams: $W$-boson loop contributions}
\label{ggfusionWboson}
\end{figure}

Charged fermions and the $W$ boson contribute to the one-loop $\gamma^*\gamma$ fusion diagrams. 
Note  that  one of the two $\gamma$'s is virtual. Since the couplings of the Higgs boson to fermions  are 
proportional to the fermion masses, we only consider  the top quark for the  charged fermion loop diagrams.
The $\gamma^*\gamma$ fusion diagrams we calculate are shown in Fig.\ref{ggfusiontquark} and \ref{ggfusionWboson}.  
The calculation is straightforward and we make full use of FeynCalc~\cite{feyncalc}. We obtain the contribution from the one-loop $\gamma^*\gamma$ fusion diagrams to the gauge-invariant scattering amplitude as follows:
\bea
A_{\gamma\gamma}&=&\Bigl(\frac{e^3 g}{16\pi^2}\Bigr)\Bigl[{\overline u}(k_1')\gamma_\mu u(k_1)\Bigr]\frac{1}{t}
\Bigl(g^{\mu\beta} -\frac{2k_2^\mu q^\beta}{m_h^2-t}\Bigr)\epsilon_\beta(k_2)~F_{\gamma\gamma}~,
\label{Agg}
\eea
with
\bea
F_{\gamma\gamma}=\frac{2 m_t^2}{m_W}N_c Q_t^2~ S^{\gamma\gamma}_{(T)}(t, m_t^2, m_h^2)
-m_W S^{\gamma\gamma}_{(W)}(t, m_W^2, m_h^2)~, \label{Fgg}
\eea
where $e$ and $g$ are the electromagnetic  and  weak gauge couplings, respectively, and $N_c=3$ and $Q_t=\frac{2}{3}$.
$S^{\gamma\gamma}_{(T)}$ and $S^{\gamma\gamma}_{(W)}$ are contributions from top loops and $W$ loops, respectively, and are expressed in terms of scalar integrals, ---more specifically, in terms of  
the Passarino-Veltman two-point integrals $B_0$'s and   three-point integrals $C_0$'s~\cite{PassarinoVeltman},
\bea
S^{\gamma\gamma}_{(T)}(t, m_t^2, m_h^2)&=&2 +  \frac{2 t}{m_h^2-t} \Bigl[B_0(m_h^2;m_t^2,m_t^2)-B_0(t;m_t^2,m_t^2)\Bigr] \nn\\
&&\hspace{4cm}+ \{ 4 m_t^2-m_h^2+t\}C_0(m_h^2,0,t;m_t^2,m_t^2,m_t^2)~,\label{SggT}\\
S^{\gamma\gamma}_{(W)}(t, m_W^2, m_h^2)&=& 6 +\frac{m_h^2-t}{m_W^2}-\frac{m_h^2 t}{2 m_W^4}\nn\\
&&+\frac{t \left(12 m_W^4+2 m_W^2 \left(m_h^2-t\right)-m_h^2 t\right)}{2
   m_W^4 \left(m_h^2-t\right)} \Bigl[B_0(m_h^2;m_W^2,m_W^2)-B_0(t;m_W^2,m_W^2)\Bigr] \nn\\
&& + \Bigl\{ \frac{t \left(m_h^2-2 t\right)}{m_W^2}+12 m_W^2-6 m_h^2+6 t\Bigr\}C_0(m_h^2,0,t;m_W^2,m_W^2,m_W^2) ~,
\label{SggW}
\eea
where $m_t$ and $m_W$ are  the top-quark and $W$-boson masses, respectively.
The explicit expressions of the relevant $B_0$'s and $C_0$'s are given in Appendix \ref{OLI}.
The two-point integrals $B_0$'s have ultraviolet divergences, but  the $B_0$'s 
in Eqs.(\ref{SggT}) and (\ref{SggW}) appear in pairs and the  differences are finite with the  ultraviolet divergences being cancelled out. The integrals $C_0$'s in Eqs.(\ref{SggT}) and.(\ref{SggW}) are finite. Therefore, $S^{\gamma\gamma}_{(T)}$ and $S^{\gamma\gamma}_{(W)}$ give  finite results.

A dimensionless quantity $G_{\gamma\gamma}(t)\equiv F_{\gamma\gamma}(t)/\Bigl(\frac{t-m_h^2}{2m_W}\Bigr)$ may be 
considered as a transition form factor of the Higgs boson. In the limit $t\rightarrow 0$, $G_{\gamma\gamma}(t)$ reduces to 
\be
G_{\gamma\gamma}(0)=N_c Q_t^2~F_{1/2}+F_1~,
\ee
where $F_{1/2}$ and $F_1$ are the top-quark and $W$-boson loop contributions to $H\rightarrow \gamma\gamma$ 
decay amplitude~\cite{H2gammas}. They are given, for example, in Eq.(2.17) of Ref.\cite{Hunter}. The $W$-boson contribution $|m_W S^{\gamma\gamma}_{(W)}|$ is much larger in magnitude than the top-quark contribution $|\frac{2 m_t^2}{m_W}N_c Q_t^2~ S^{\gamma\gamma}_{(T)}|$ and grows with $-t$. Thus, $G_{\gamma\gamma}(t)$, the sum of 
top-quark and $W$-boson contributions, grows with $-t$. Actually, it grows as $\log^2 \frac{-t}{m_W^2}$ for large $-t$~.

\subsection{$Z$ boson--real photon fusion diagrams}

The one-loop $Z^*\gamma$ fusion diagrams for the Higgs boson production are obtained from the one-loop 
$\gamma^*\gamma$ fusion diagrams given in Figs.\ref{ggfusiontquark} and \ref{ggfusionWboson} by replacing the photon propagator with that of the
 $Z$ boson with mass $m_Z$.
Charged fermions and $W$ boson contribute to the one-loop $Z^*\gamma$ fusion diagrams. 
Again we only consider the top quark for the  charged fermion loop diagrams.
We calculate the contribution from the $Z^*\gamma$ fusion diagrams  and obtain,
\bea
A_{Z\gamma}&=&\Bigl(\frac{e g^3}{16\pi^2}\Bigr)\biggl[{\overline u}(k_1')\gamma_\mu\Bigl(f_{Ze}+\gamma_5\Bigr)u(k_1)\biggr]\frac{1}{t-m_Z^2}
\Bigl(g^{\mu\beta} -\frac{2k_2^\mu q^\beta}{m_h^2-t}\Bigr)\epsilon_\beta(k_2)~F_{Z\gamma} ~,
\label{AZg}
\eea
with
\bea
F_{Z\gamma}=-\frac{ m_t^2}{8m_W \cos^2\theta_W}N_c Q_t f_{Zt}S^{Z\gamma}_{(T)}(t, m_t^2, m_h^2)
+\frac{m_W}{4} S^{Z\gamma}_{(W)}(t, m_W^2, m_h^2)~,  \label{FZg}
\eea
where $f_{Ze}$ and $f_{Zt}$ are the strength of the vector part of the $Z$-boson coupling to the
electron and top quark, respectively, and are given by
\bea
f_{Ze}=-1+4\sin^2\theta_W~,\qquad f_{Zt}=1-\frac{8}{3}\sin^2\theta_W~,
\eea
with $\theta_W$ being the Weinberg angle.
The axial-vector part of the $Z$-boson coupling to the top quark [see Eq.(\ref{ttZcoupling})] has a null  
effect and we find 
\bea
S^{Z\gamma}_{(T)}(t, m_t^2, m_h^2)=S^{\gamma\gamma}_{(T)}(t, m_t^2, m_h^2)~,\qquad 
S^{Z\gamma}_{(W)}(t, m_W^2, m_h^2)=S^{\gamma\gamma}_{(W)}(t, m_W^2, m_h^2)~.
\eea

\subsection{``$W\nu_e$" one-loop diagrams}

\begin{figure}[htbp]
 \begin{tabular}{cc}
 \begin{minipage}{0.33\hsize}
  \begin{center}
   \includegraphics[width=50mm]{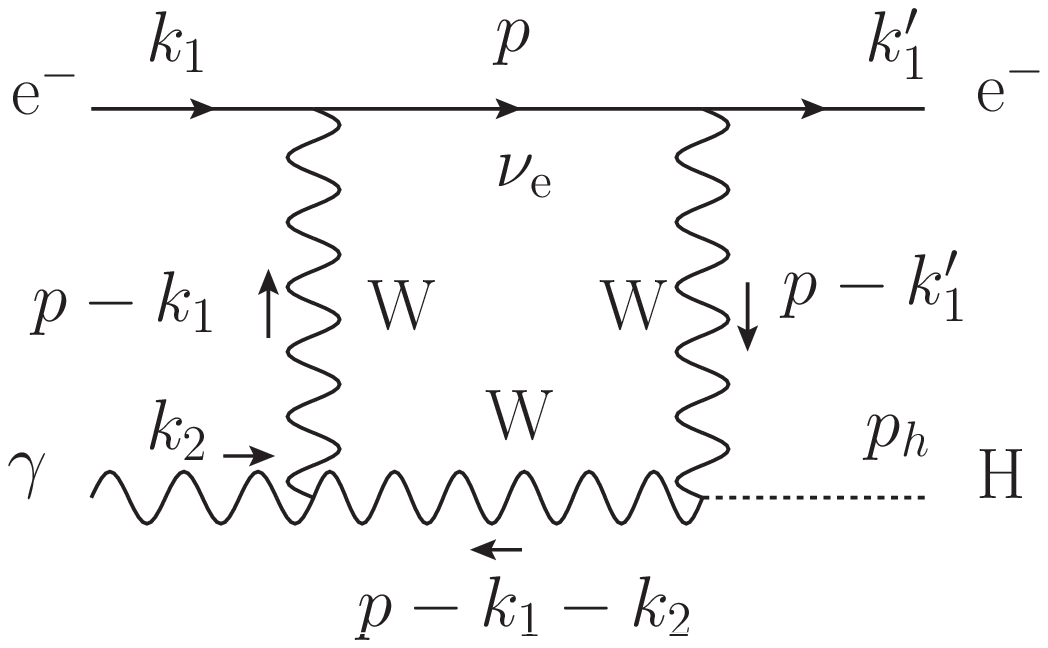}
  \end{center}
 \end{minipage} &
 \begin{minipage}{0.33\hsize}
 \begin{center}
  \includegraphics[width=50mm]{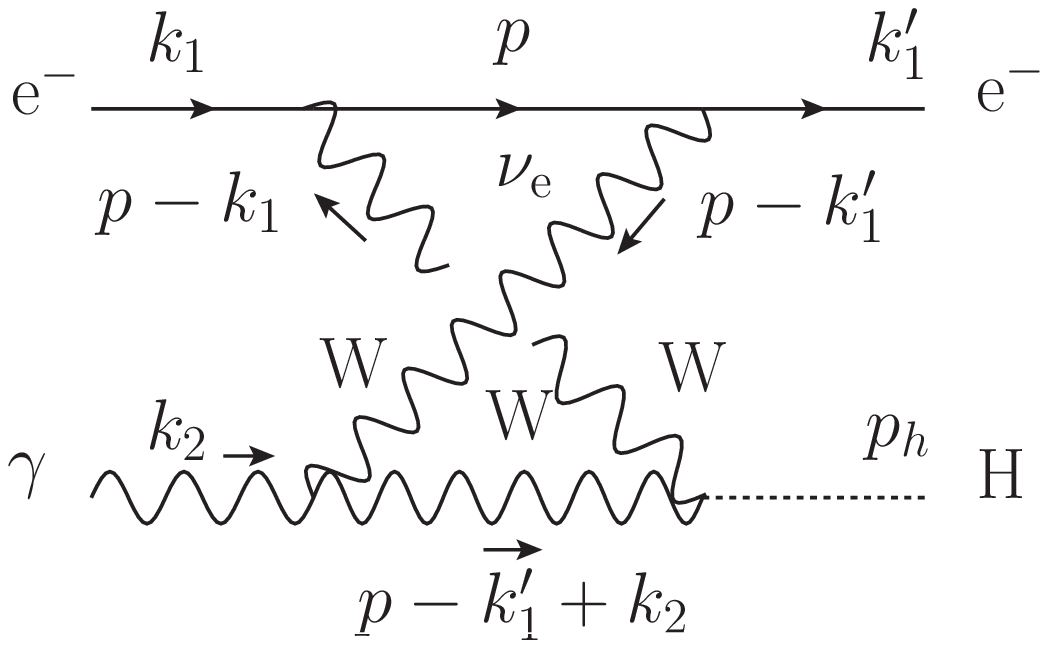}
 \end{center}
 \end{minipage}\\
 \begin{minipage}{0.33\hsize}
 \begin{center}
  \includegraphics[width=50mm]{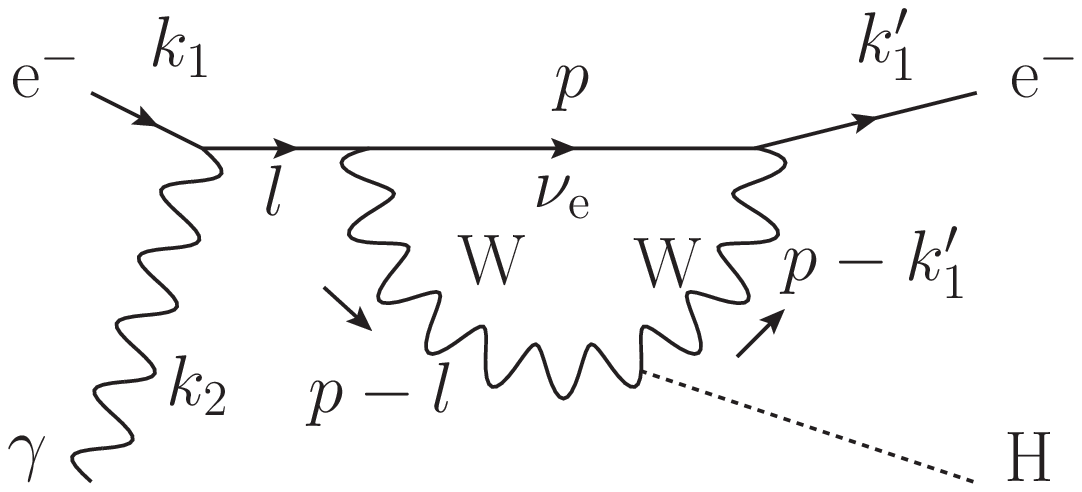}
 \end{center}
 \end{minipage}  &
\begin{minipage}{0.33\hsize}
 \begin{center}
  \includegraphics[width=50mm]{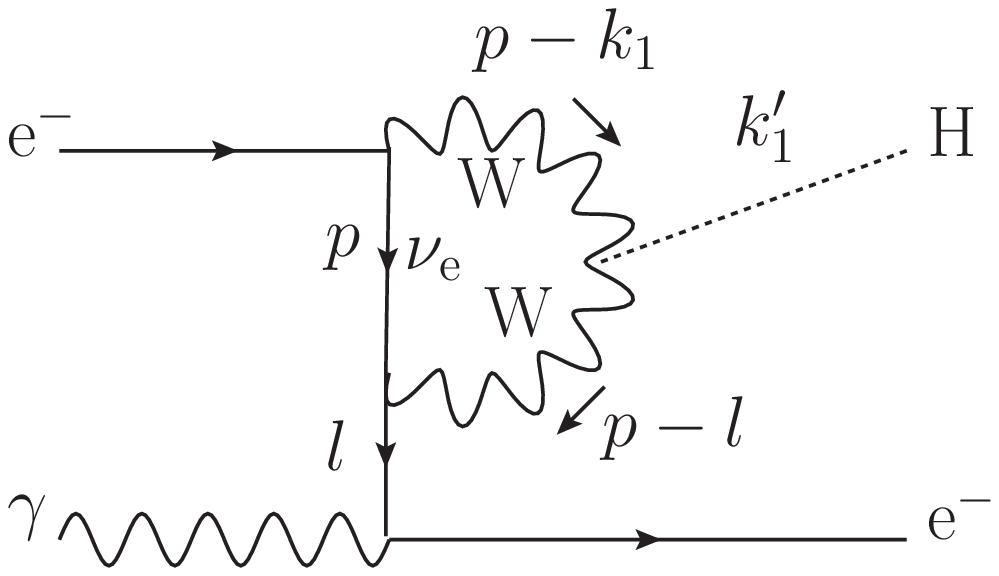}
 \end{center}
  \label{fig:three}
 \end{minipage}
 \end{tabular}
 \caption{ ``$W\nu_e$" diagrams}
 \label{Wrelated}
\end{figure}

The Feynman diagrams involving the $W$ boson and electron neutrino, which are shown in Fig.\ref{Wrelated}, also contribute to the Higgs boson production in $e^-\gamma$ collisions. They yield the ``$W\nu_e$" amplitude which is written 
in the following form,
\bea
A_{W\nu_e}=\Bigl(\frac{eg^3}{16\pi^2}\Bigr)\frac{ m_W}{4} \Bigl[{\overline u}(k_1')~F_{(W\nu_e)\beta}~(1-\gamma_5 )u(k_1)\Bigr]\epsilon(k_2)^\beta~,
\eea
where the factor $(1-\gamma_5)$ is due to the $e$-$\nu$-$W$ vertex. Thus, when the  electron beams are right-handedly polarized, these ``$W\nu_e$" diagrams do not contribute. 
The factor $F_{(W\nu_e)\beta}$ is written in a gauge-invariant form as
\bea
F_{(W\nu_e)\beta}&=&\Bigl(\frac{2k_{1_\beta} \ksl_2}{s}-\gamma_\beta \Bigr)S^{{W\nu_e}}_{(k_{1})}(s,t,m_h^2,m_W^2)+\Bigl(\frac{2k'_{1_\beta} \ksl_2}{u}+\gamma_\beta\Bigr)  S^{{W\nu_e}}_{(k'_{1})}(s,t,m_h^2,m_W^2)~,\label{FWnu}
\eea
where $S^{{W\nu_e}}_{(k_{1})}$ and $S^{{W\nu_e}}_{(k'_{1})}$ are expressed in terms of  the scalar 
integrals $B_0$'s, $C_0$'s and the scalar four-point integrals $D_0$'s as follows: 
\bea
&&S^{{W\nu_e}}_{(k_{1})}(s,t,m_h^2,m_W^2)\nn\\
&&\qquad =  \frac{ (2m_W^2+m_h^2)s}{2m_W^4 (s+u)} +\frac{2}{(s+ t)}\Bigl[B_0\left(m_h^2;m_W^2,m_W^2\right)-B_0(u;0,m_W^2)\Bigr] \nn\\
&&\qquad\quad+\frac{st \left(2 m_W^2+m_h^2\right)}{2m_W^4 (s+u)^2}\Bigl[B_0\left(m_h^2;m_W^2,m_W^2\right)-B_0\left(t;m_W^2,m_W^2\right)\Bigr]\nn\\
&&\qquad\quad +  \frac{ \left(m_W^2-s\right)}{ t}C_0\left(0,0,s;m_W^2,m_W^2,0\right) -\frac{ u \left(-m_W^2+s+t\right)}{s t}C_0\left(0,0,u;m_W^2,m_W^2,0\right)\nn\\
&&\qquad\quad  -\frac{t}{s}C_0\left(0,0,t;m_W^2,0,m_W^2\right)+  \frac{ \left(m_W^2-s\right) (t+u)}{s t}C_0\left(0,s,m_h^2;m_W^2,0,m_W^2\right)\nn\\
&&\qquad\quad  -\frac{ \left(s^2-2 s t-t^2\right) \left(-m_W^2+s+t\right)}{s t (s+t)}C_0\left(0,u,m_h^2;m_W^2,0,m_W^2\right)\nn\\
&&\qquad\quad +  \frac{-2 m_W^4 (s+u)^2+ m_W^2 \left(2 s^3+s^2 (3 t+4 u)+2 s u (t+u)+t u^2\right)- s^2 t(s-t+u)}{m_W^2 s t (s+u)}\nn\\
&&\hspace{7cm}\times C_0\left(0,t,m_h^2;m_W^2,m_W^2,m_W^2\right)\nn\\
&&\qquad\quad   + \frac{ \left(m_W^2-s\right) \left(m_W^2 (s+u)+s t\right)}{s t}
D_0\left(0,0,0,m_h^2;s,t;m_W^2,m_W^2,0,m_W^2\right)\nn\\
&&\qquad\quad  + \frac{ \left(m_W^4 (s+u)-m_W^2 \left(s^2+s (u-t)+2 t u\right)+t u (s+t)\right)}{s t}
\nn\\
&&\hspace{7cm}\times D_0\left(0,0,0,m_h^2;t,u;m_W^2,0,m_W^2,m_W^2\right)~,\label{SWk1b}
\eea
and 
\bea
&&S^{{W\nu_e}}_{(k'_{1})}(s,t,m_h^2,m_W^2)\nn\\
&&\qquad =  -\frac{(2 m_W^2+m_h^2)u}{2m_W^4 (s+u)}  - \frac{2}{(t+u)}\Bigl[B_0\left(m_h^2;m_W^2,m_W^2\right)-B_0(s;0,m_W^2)\Bigr] \nn\\
&&\qquad \quad -\frac{t u \left(2 m_W^2+m_h^2\right)}{2m_W^4 (s+u)^2}\Bigl[B_0\left(m_h^2;m_W^2,m_W^2\right)-B_0\left(t;m_W^2,m_W^2\right)\Bigr]\nn\\
&&\qquad\quad +  \frac{ s \left(-m_W^2+t+u\right)}{t u}C_0\left(0,0,s;m_W^2,m_W^2,0\right)  -\frac{\left(m_W^2-u\right)}{t }C_0\left(0,0,u;m_W^2,m_W^2,0\right)\nn\\
&&\qquad\quad  + \frac{ t}{u}C_0\left(0,0,t;m_W^2,0,m_W^2\right) +\frac{ \left(u^2-2tu-t^2\right) \left(-m_W^2+t+u\right)}{t u (t+u)}C_0\left(0,s,m_h^2;m_W^2,0,m_W^2\right)\nn\\
&&\qquad\quad   -\frac{ \left(m_W^2-u\right) (s+t)}{t u}C_0\left(0,u,m_h^2;m_W^2,0,m_W^2\right)\nn\\
&&\qquad\quad +   \frac{2 m_W^4 (s+u)^2-m_W^2 \left(2u^3+u^2(3t+4s)+2su(s+t)+s^2 t\right)+t
   u^2 (s-t+u)}{m_W^2 t u (s+u)}\nn\\
&&\hspace{7cm}\times C_0\left(0,t,m_h^2;m_W^2,m_W^2,m_W^2\right)\nn\\
&&\qquad\quad   -\frac{ \left(m_W^2-u\right) \left(m_W^2 (s+u)+t u\right)}{t u}
D_0\left(0,0,0,m_h^2;t,u;m_W^2,0,m_W^2,m_W^2\right)\nn\\
&&\qquad\quad    -\frac{ \left(m_W^4 (s+u)-m_W^2 (s (2 t+u)+u (u-t))+s t (t+u)\right)}{t u}
\nn\\
&&\hspace{7cm}\times
D_0\left(0,0,0,m_h^2;s,t;m_W^2,m_W^2,0,m_W^2\right)~.\label{SWk1bdash}
\eea
The explicit expressions of $B_0$'s, $C_0$'s and $D_0$'s are given in 
Appendix \ref{OLI}.
The integrals  $C_0$'s and $D_0$'s in Eqs.(\ref{SWk1b}) and (\ref{SWk1bdash}) are all finite.
Again, the integrals $B_0$'s  appear in pairs and the  differences  yield  finite results. In the end, $S^{{W\nu_e}}_{(k_{1})}$ and $S^{{W\nu_e}}_{(k'_{1})}$ are finite. Finally we note that $S^{{W\nu_e}}_{(k'_{1})}$ vanishes at $u=0$, which is anticipated from the expression of the second term in Eq.(\ref{FWnu}).

\subsection{``$Ze$" one-loop diagrams}

\begin{figure}[htbp]
 \begin{tabular}{c}
 \begin{minipage}{0.33\hsize}
  \begin{center}
   \includegraphics[width=50mm]{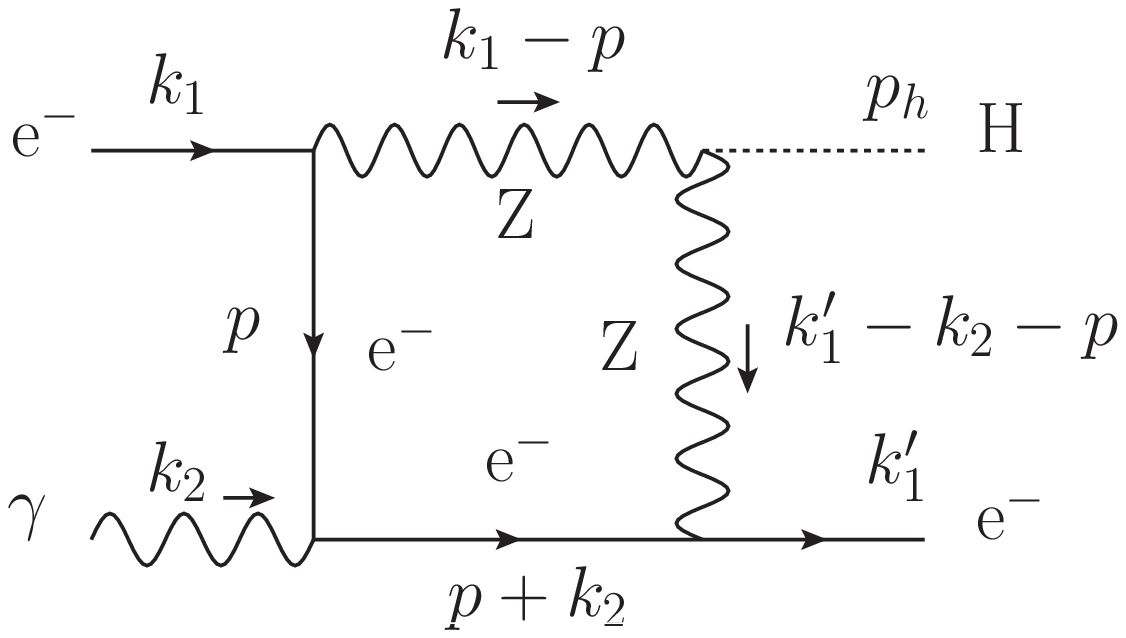}
  \end{center}
 \end{minipage}
 \begin{minipage}{0.33\hsize}
 \begin{center}
  \includegraphics[width=50mm]{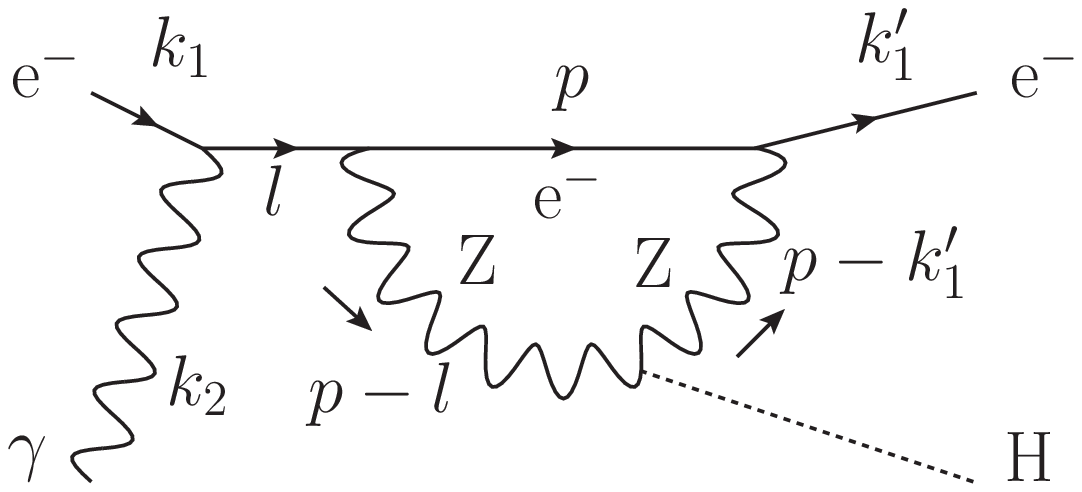}
 \end{center}
 \end{minipage}
 \begin{minipage}{0.33\hsize}
 \begin{center}
  \includegraphics[width=50mm]{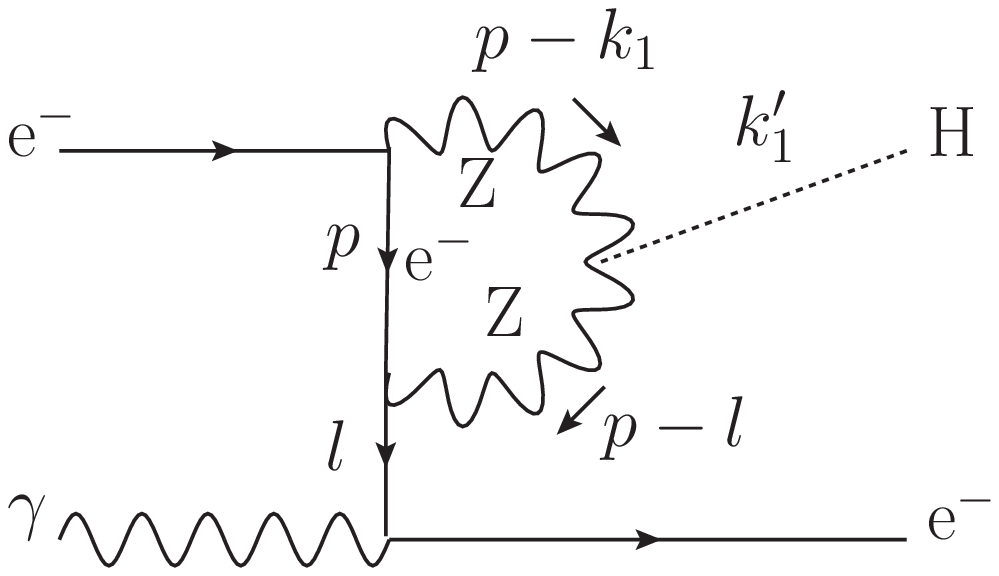}
 \end{center}
 \end{minipage}
 \end{tabular}
 \caption{ ``$Ze$" diagrams}
 \label{Zrelated} 
\end{figure}

The last one-loop contributions to the Higgs boson production in $e^-\gamma$ collisions come  from the Feynman diagrams shown in Fig.\ref{Zrelated}. These ``$Ze$" diagrams give the following amplitude,
\bea
A_{Ze}=\Bigl(\frac{eg^3}{16\pi^2}\Bigr)\Bigl(-\frac{ m_Z}{16\cos^3\theta_W}\Bigr)\times \Bigl[{\overline u}(k_1')~F_{(Ze)\beta}~(f_{Ze}+\gamma_5 )^2u(k_1)\Bigr]\epsilon(k_2)^\beta~,
\eea
where the factor $(f_{Ze}+\gamma_5 )^2$ arises from the $Z$-boson coupling to electrons. 
The factor $F_{(Ze)\beta}$ is written in a gauge-invariant form as
\bea
F_{(Ze)\beta}
&=&\Bigl(\frac{2k_{1_\beta} \ksl_2}{s}-\gamma_\beta \Bigr) S^{Ze}_{(k_{1})}(s,t,m_h^2,m_Z^2)+\Bigl(\frac{2k'_{1_\beta} \ksl_2}{u}+\gamma_\beta\Bigr)  S^{Ze}_{(k'_{1})}(s,t,m_h^2,m_Z^2)~,
\eea
where
\bea
&&S^{Ze}_{(k_{1})}(s,t,m_h^2,m_Z^2)\nn\\
&&\qquad = -\frac{2}{ (s+t)}\Bigl[B_0\left(m_h^2;m_Z^2,m_Z^2\right) -B_0\left(u;0,m_Z^2\right)\Bigr]  -\frac{ \left(m_Z^2-s\right) (t+u)}{s t}
C_0\left(0,s,m_h^2;m_Z^2,0,m_Z^2\right)\nn\\
&& \qquad\quad+ \frac{ \left(m_Z^2 \left(-s^2+2 s t+t^2\right)+s^3-s t^2\right)}{s t (s+t)}C_0\left(0,u,m_h^2;m_Z^2,0,m_Z^2\right) \nn\\
&&\qquad\quad+  \frac{ \left(m_Z^2-s\right)}{s t}\Bigl\{sC_0\left(0,0,s;m_Z^2,0,0\right)+u
C_0\left(0,0,u;m_Z^2,0,0\right)\nn\\
&&\hspace{4cm}+\Bigl[m_Z^2 (s+u)-s u\Bigr]D_0\left(0,0,0,m_h^2;s,u;m_Z^2,0,0,m_Z^2\right)
\Bigr\}~,\label{SZk1b}
\eea
and 
\bea
&&S^{Ze}_{(k'_{1})}(s,t,m_h^2,m_Z^2)\nn\\
&&\qquad =  \frac{2}{ (t+u)}\Bigl[B_0\left(m_h^2;m_Z^2,m_Z^2\right)-B_0\left(s;0,m_Z^2\right) \Bigr]  + \frac{ \left(m_Z^2-u\right) (s+t)}{t u}C_0\left(0,u,m_h^2;m_Z^2,0,m_Z^2\right) \nn\\
&&\qquad\quad  -\frac{ \left(m_Z^2 \left(t^2+2 t u-u^2\right)-t^2 u+u^3\right)}{t u (t+u)}
C_0\left(0,s,m_h^2;m_Z^2,0,m_Z^2\right)\nn\\
&&\qquad\quad -  \frac{ \left(m_Z^2-u\right)}{ t u}\Bigl\{sC_0\left(0,0,s;m_Z^2,0,0\right)+u
C_0\left(0,0,u;m_Z^2,0,0\right)\nn\\
&&\hspace{4cm}+\Bigl[m_Z^2 (s+u)-s u\Bigr]D_0\left(0,0,0,m_h^2;s,u;m_Z^2,0,0,m_Z^2\right)
\Bigr\}~.\label{SZk1bdash}
\eea 
The explicit expressions of the scalar integrals $B_0$'s, $C_0$'s and $D_0$'s  which appear in Eqs.(\ref{SZk1b}) and (\ref{SZk1bdash})  are given in Appendix \ref{OLI}.
The  integrals $C_0\left(0,s,m_h^2;m_Z^2,0,m_Z^2\right)$ and $C_0\left(0,u,m_h^2;m_Z^2,0,m_Z^2\right)$  are finite. On the other hand, 
collinear singularities appear in $C_0\left(0,0,s;m_Z^2,0,0\right)$, $C_0\left(0,0,u;m_Z^2,0,0\right)$ and in the four-point integral $D_0\left(0,0,0,m_h^2;s,u;m_Z^2,0,0,m_Z^2\right)$. These collinear divergences 
are handled by dimensional regularization. See Eqs.(\ref{C0Sings}), (\ref{C0Singu}) and (\ref{D0Sing}). These scalar integrals with collinear divergences appear in combination 
as in the parentheses of the last terms of Eqs.(\ref{SZk1b}) and (\ref{SZk1bdash}) and, as a result, their collinear divergences cancel out. Thus $S^{Ze}_{(k_{1})}$ and $S^{Ze}_{(k'_{1})}$ are both finite. Note also that $S^{Ze}_{(k'_{1})}$ vanishes at $u=0$.

\section{Higgs boson production cross section \label{Section3}}

One of the advantages of  linear colliders is that we can acquire highly
polarized colliding beams. Let us consider the Higgs boson production 
reaction (\ref{HiggsProduction}) when both the initial electron and photon beams 
are fully polarized. We denote the polarizations of the electron and photon  as $P_e=\pm 1$ and $P_\gamma=\pm 1$, respectively\footnote{For the cases of full polarization, we note $P_e=2\times{\rm electron~ helicity}$ and $P_\gamma=$ photon helicity.}. The differential cross section for $e^-\gamma\rightarrow e^-H$ with the initial electron and photon polarizations $P_e$ and $P_\gamma$ is expressed by,
\bea
\frac{d\sigma_{(e\gamma\rightarrow eH)}(s,P_e,P_\gamma)}{dt}
&=&\frac{1}{16\pi s^2}
\times \biggl\{\sum_{\rm final~electron~spin}~|A(P_e,P_\gamma)|^2\biggr\}~,\label{CrossSectionFormula}
\eea
where $A(P_e, P_\gamma)$ is written at the one-loop level as 
\bea
A(P_e, P_\gamma)&=&A_{\gamma\gamma}(P_e, P_\gamma)+A_{Z\gamma}(P_e, P_\gamma)+A_{W\nu_e}(P_e, P_\gamma)+A_{Ze}(P_e, P_\gamma)~.  \label{FourAmplitude}
\eea
In the center-of-mass (CM) frame, $t$ and $u$ are expressed as
\bea
t=-\frac{s-m_h^2}{2}(1-\cos\theta)~,\qquad \qquad u=-\frac{s-m_h^2}{2}(1+\cos\theta)~, \label{CMangle}
\eea
where $\theta$ is the angle between the initial and scattered electrons. We are dealing 
with   $e^-\gamma$ collisions in  the high-energy limit and thus we neglect the electron mass. In the massless limit the helicity of the electron is conserved. Then the angular momentum conservation along the direction of the initial 
electron requires that the amplitude $A(P_e, P_\gamma)$  should vanish at $\theta=0$. Hence, apart from the photon 
propagator which appears as $\frac{1}{t}$, an overall factor $t$ arises in the  differential cross section. 
Also when the  electron is scattered in the backward direction, the amplitude  $A(P_e, P_\gamma)$ with $P_eP_\gamma=-1$ should vanish at $\theta=\pi$ due to the angular momentum conservation.  Hence the differential cross section for the initial beams with $P_eP_\gamma=-1$ vanishes as $u\rightarrow 0$ (or $t\rightarrow t_{\rm min}=m_h^2-s$).

When an initial electron is polarized with polarization $P_e$, we modify $u(k_1)$  as
\bea
u(k_1)\rightarrow \frac{1+P_e \gamma_5}{2}~u(k_1)~.\label{Projection}
\eea
In the center-of-mass frame where a photon with momentum $k_2$ is moving in the $+z$ direction, the circular polarization ($P_\gamma=\pm1$) of  the photon is taken to be  
\bea
\epsilon(k_2, \pm 1)_\beta=\frac{1}{\sqrt{2}}(0, \mp 1, -i,0)~. \label{CircularPolarization}
\eea
In this frame, the momenta $k_2$, $k_1$ and $k'_1$ are expressed as 
\bea
k_2^\mu=\frac{\sqrt{s}}{2}(1,0,0,1),\quad k_1^\mu=\frac{\sqrt{s}}{2}(1,0,0,-1), \quad
{k'}_1^\mu =\frac{s-m_h^2}{\sqrt{s}}(1, \sin\theta \cos\phi, \sin\theta \sin\phi, -\cos\theta) ~,\label{SpecialFrame}
\eea
and  we find that the polarization tensor of the
circularly polarized photon is given by 
\bea
\epsilon(k_2, \pm 1)^*_\alpha \epsilon(k_2, \pm 1)_\beta=-\frac{1}{2}~g_{\alpha\beta}\pm \frac{i}{2}~\Bigl(
g_{\alpha 1}g_{\beta 2}-g_{\alpha 2}g_{\beta 1}\Bigr)~.
\eea
With $\varepsilon^{0123}=1$, we obtain in this frame
\bea
\Bigl\{(k'_1)^1 \varepsilon^{2\mu\nu\lambda}-(k'_1)^2 \varepsilon^{1\mu\nu\lambda}\Bigr\}k_{1\mu} k_{2\nu} k'_{1\lambda}=-\frac{1}{2} tu~,\qquad \qquad 
\varepsilon^{12\mu\nu}k_{1\mu} k'_{1\nu}=\frac{1}{2}t~.\label{relations}
\eea

Using Eqs.(\ref{Projection}) - (\ref{relations}), we evaluate $\sum_{\rm final~electron~spin}~|A(P_e,P_\gamma)|^2$ and obtain the differential cross section for $e^-\gamma\rightarrow e^-H$ for each case of  polarizations of the  electron and photon beams.
In order to see the relative contributions from  $\gamma^*\gamma$ fusion, 
$Z^*\gamma$ fusion,  and the ``$W\nu_e$"  and  ``$Ze$"  diagrams,  we evaluate the differential cross section given in Eq.(\ref{CrossSectionFormula}) by replacing $A(P_e, P_\gamma)$ with $A_{\gamma\gamma}(P_e, P_\gamma)$, $A_{Z\gamma}(P_e, P_\gamma)$, $A_{W\nu_e}(P_e, P_\gamma)$ and $A_{Ze}(P_e, P_\gamma)$, respectively. We obtain
\bea
\frac{d\sigma_{(\gamma\gamma)}(s,P_e,P_\gamma)}{dt}
&=&\frac{1}{16\pi~s^2}\Bigl(\frac{e^3g}{16\pi^2}\Bigr)^2 \left(-\frac{1}{t}\right){F_{\gamma\gamma}}^2~\Bigl\{ \frac{s^2+u^2}{(s+u)^2}+P_\gamma P_e \Bigl(1-\frac{2u}{s+u}\Bigr) \Bigr\},\label{dsigmadtgammagamma}\\
\frac{d\sigma_{(Z\gamma)}(s,P_e,P_\gamma)}{dt}
&=&\frac{1}{16\pi~s^2}\Bigl(\frac{eg^3}{16\pi^2}\Bigr)^2 \frac{-t}{(t-m_Z^2)^2}~{F_{Z\gamma}}^2~
\nn\\
&&\times\Bigl\{ (f^2_{Ze}+2 P_e f_{Ze}+1)\frac{s^2+u^2}{(s+u)^2}+P_\gamma(P_ef^2_{Ze}+2  f_{Ze}+P_e) \Bigl(1-\frac{2u}{s+u}\Bigr) \Bigr\}, \label{dsigmadtZgamma}\\
\frac{d\sigma_{(W\nu_e)}(s,P_e,P_\gamma)}{dt}
&=&\frac{1}{16\pi s^2}\Bigl(\frac{eg^3}{16\pi^2}\Bigr)^2 \frac{ m_W^2}{8}(-t)(1-P_e)\Biggl\{ \biggl[  \Big|S^{{W\nu_e}}_{(k_{1})}(s,t,m_h^2,m_W^2)\Big|^2+ \Big|S^{{W\nu_e}}_{(k'_{1})}(s,t,m_h^2,m_W^2)\Big|^2  \biggr]\nn\\
&&+P_\gamma\biggl[- \Big|S^{{W\nu_e}}_{(k_{1})}(s,t,m_h^2,m_W^2)\Big|^2+ \Big|S^{{W\nu_e}}_{(k'_{1})}(s,t,m_h^2,m_W^2)\Big|^2 \biggr]\Biggr\},\label{dsigmadtWnu}\\
\frac{d\sigma_{(Ze)}(s,P_e,P_\gamma)}{dt}
&=&\frac{1}{16\pi s^2}\Bigl(\frac{eg^3}{16\pi^2}\Bigr)^2 \Bigl(\frac{m_Z}{16\cos^3\theta_W}\Bigr)^2(-t)\nn\\
&&{\hspace{-1cm}}\times\Biggl\{ (f_{Ze}^4+4P_e f_{Ze}^3+6f_{Ze}^2+4P_ef_{Ze}+1)\biggl[  \Big|S^{Ze}_{(k_{1})}(s,t,m_h^2,m_Z^2)\Big|^2+ \Big|S^{Ze}_{(k'_{1})}(s,t,m_h^2,m_Z^2)\Big|^2  \biggr]\nn\\
&& \quad +P_\gamma(P_ef_{Ze}^4+4 f_{Ze}^3+6P_ef_{Ze}^2+4f_{Ze}+P_e)\biggl[ \Big|S^{Ze}_{(k_{1})}(s,t,m_h^2,m_Z^2)\Big|^2\nn\\
&&\hspace{8cm} - \Big|S^{Ze}_{(k'_{1})}(s,t,m_h^2,m_Z^2)\Big|^2 \biggr]  \Biggr\}~.\label{dsigmadtZe}
\eea
When the initial electron is right-handed ($P_e=+1$), there is no contribution from the ``$W\nu_e$" diagrams and we indeed see that~ $d\sigma(s,P_e=+1,P_\gamma)_{(W\nu_e)}/dt=0$. Also using the fact that 
$S^{{W\nu_e}}_{(k'_{1})}$ and $S^{Ze}_{(k'_{1})}$ vanish at $u=0$, we find that the above four differential cross sections reduce to zero as $u\rightarrow 0$ for the case $P_eP_\gamma=-1$.
In order to examine the differential cross section for $e^-\gamma\rightarrow e^- H$ given in Eq.(\ref{CrossSectionFormula}), we need to evaluate the interference terms among the four amplitudes $A_{\gamma\gamma}(P_e, P_\gamma)$, $A_{Z\gamma}(P_e, P_\gamma)$, $A_{W\nu_e}(P_e, P_\gamma)$ and $A_{Ze}(P_e, P_\gamma)$. The expressions of the six interference terms are given in Appendix \ref{Interference}.

\begin{figure}[htbp]
 \begin{tabular}{cc}
 \begin{minipage}{0.45\hsize}
  \begin{center}
   \includegraphics[width=80mm]{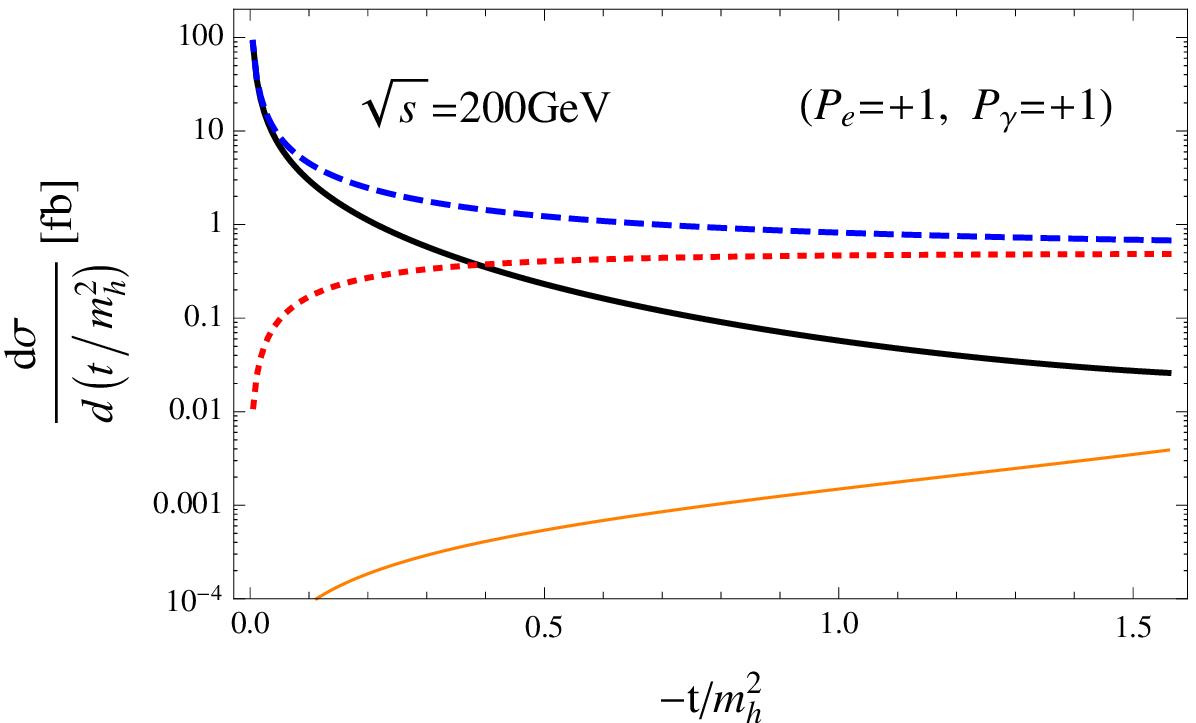}
  \end{center}
 \end{minipage} &
 \begin{minipage}{0.45\hsize}
 \begin{center}
  \includegraphics[width=80mm]{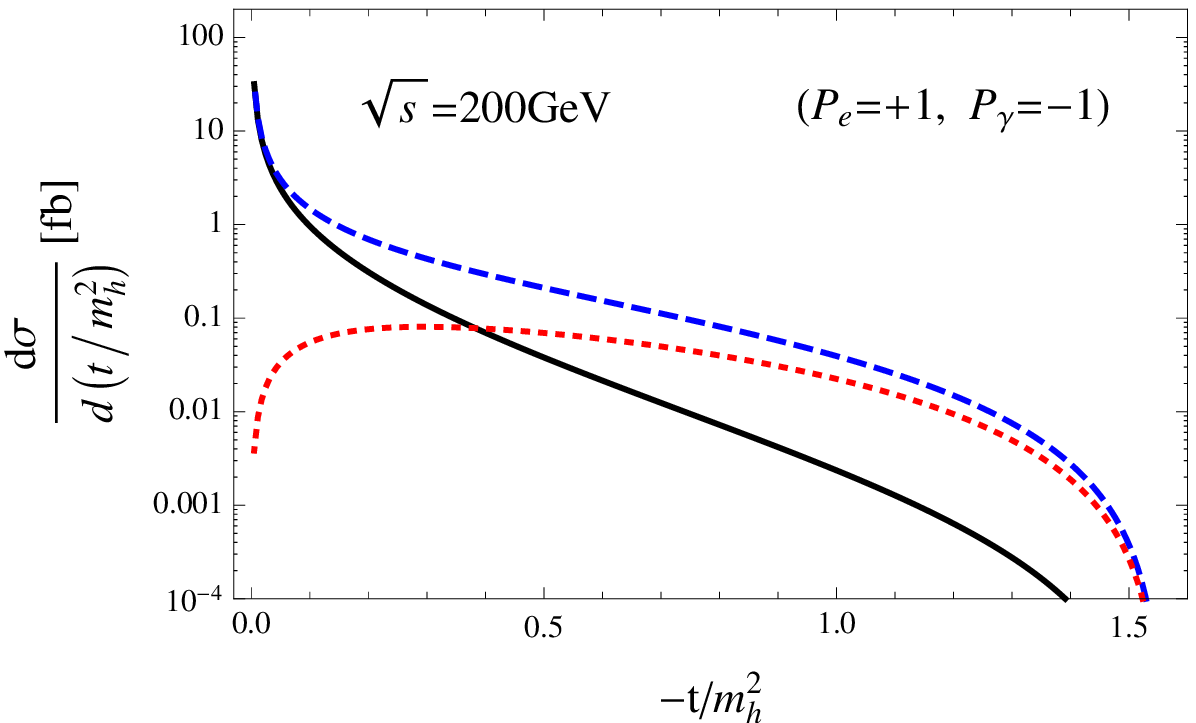}
 \end{center}
 \end{minipage}\\
 \begin{minipage}{0.45\hsize}
 \begin{center}
  \includegraphics[width=80mm]{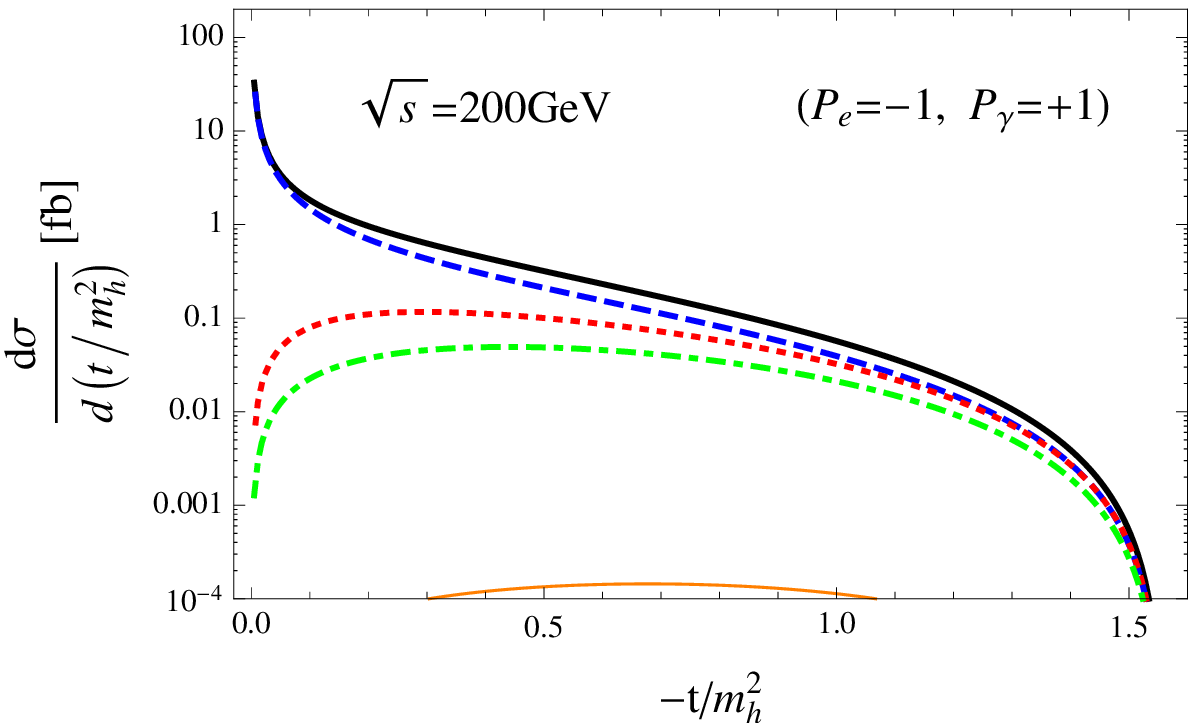}
 \end{center}
 \end{minipage}  &
\begin{minipage}{0.45\hsize}
 \begin{center}
  \includegraphics[width=80mm]{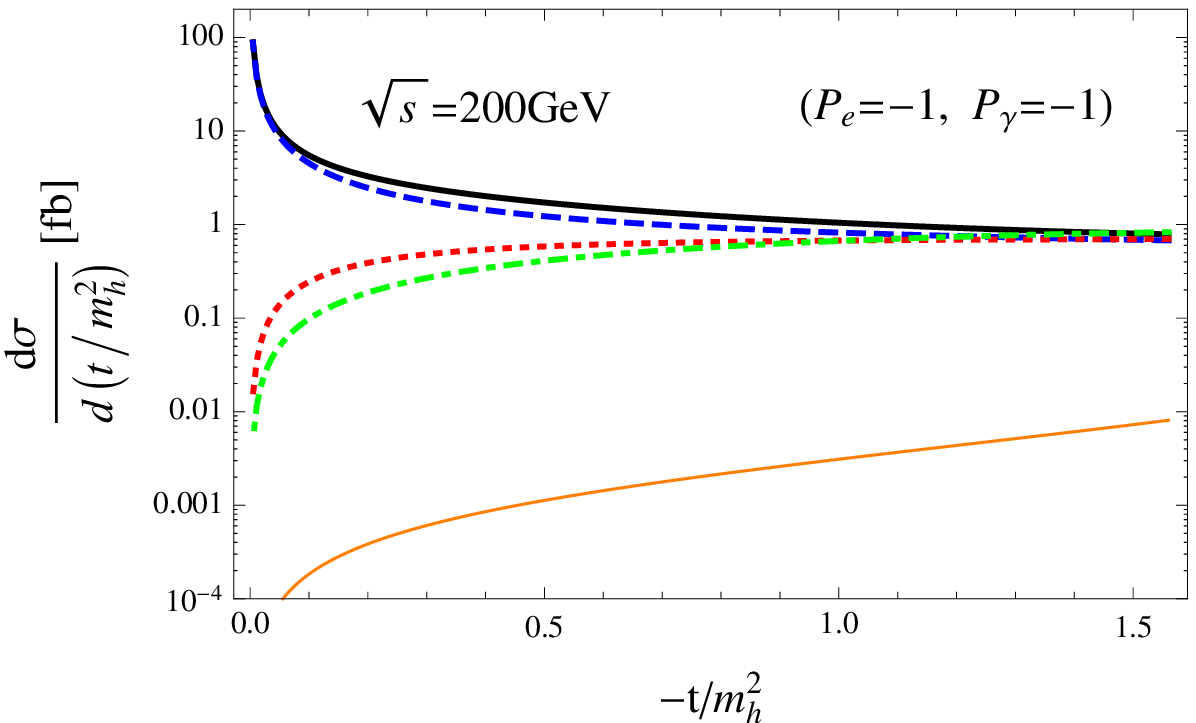}
 \end{center}
  \label{fig:three}
 \end{minipage}
 \end{tabular}
 \caption{The differential cross section for Higgs boson production $d\sigma_{(e\gamma\rightarrow eH)}/d(t/m_h^2)$  (black solid line) together with  $d\sigma_{(\gamma\gamma)}/d(t/m_h^2)$  (blue dashed line), $d\sigma_{(Z\gamma)}/dt$  (red dotted line), $d\sigma_{(W\nu_e)}/d(t/m_h^2)$  (green dot-dashed line) and $d\sigma_{(Ze)}/d(t/m_h^2)$  (orange thin solid line) as a function of
$-t/m_h^2$ with $\sqrt{s}=200$ GeV for four cases of  polarizations of the initial electron and photon beams, $(P_e=+1,P_\gamma=+1)$, $(P_e=+1,P_\gamma=-1)$, $(P_e=-1,P_\gamma=+1)$ and $(P_e=-1,P_\gamma=-1)$. In the plot of $(P_e=+1,P_\gamma=-1)$, $d\sigma_{(Ze)}/dt$ is too small and is out of the plot range.}
 \label{dsigmadt200}
\end{figure}

\begin{figure}[htbp]
 \begin{tabular}{cc}
 \begin{minipage}{0.45\hsize}
  \begin{center}
   \includegraphics[width=80mm]{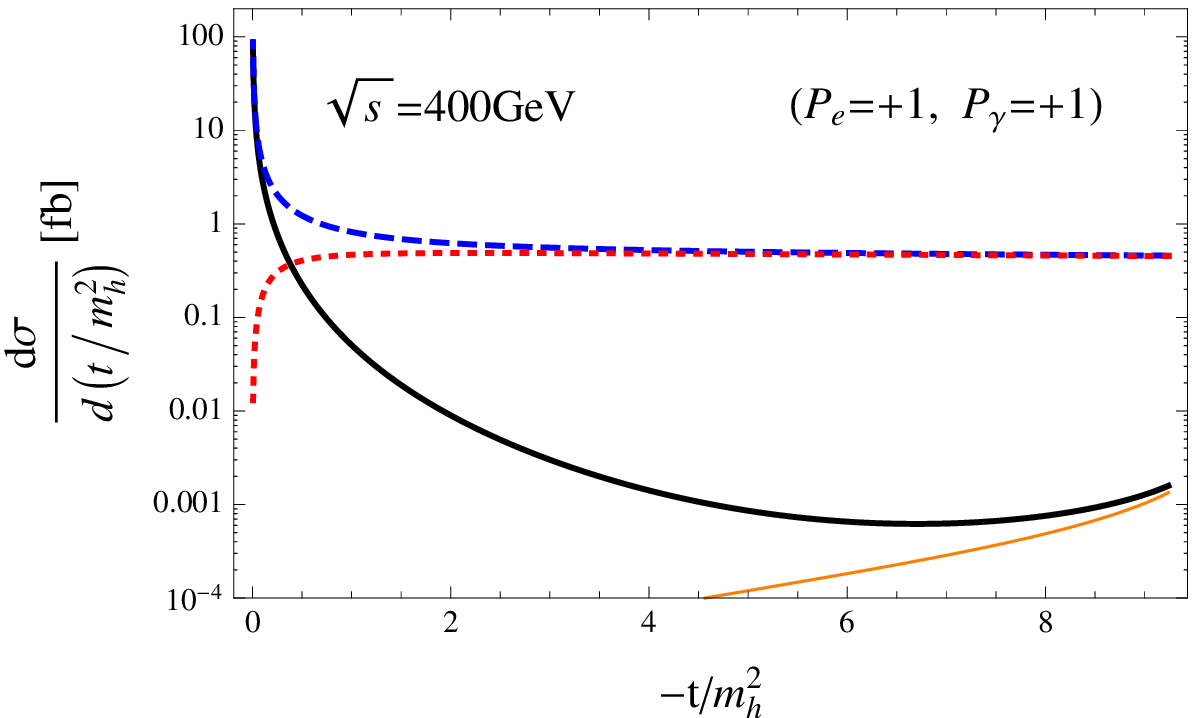}
  \end{center}
 \end{minipage} &
 \begin{minipage}{0.45\hsize}
 \begin{center}
  \includegraphics[width=80mm]{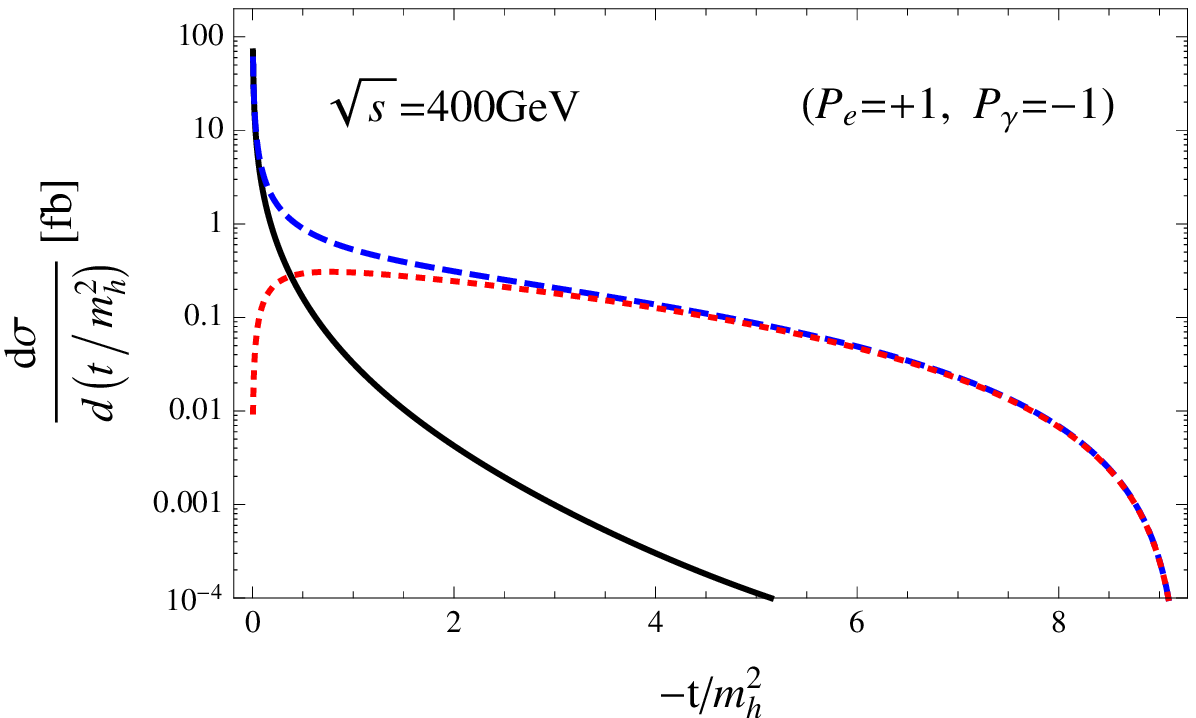}
 \end{center}
 \end{minipage}\\
 \begin{minipage}{0.45\hsize}
 \begin{center}
  \includegraphics[width=80mm]{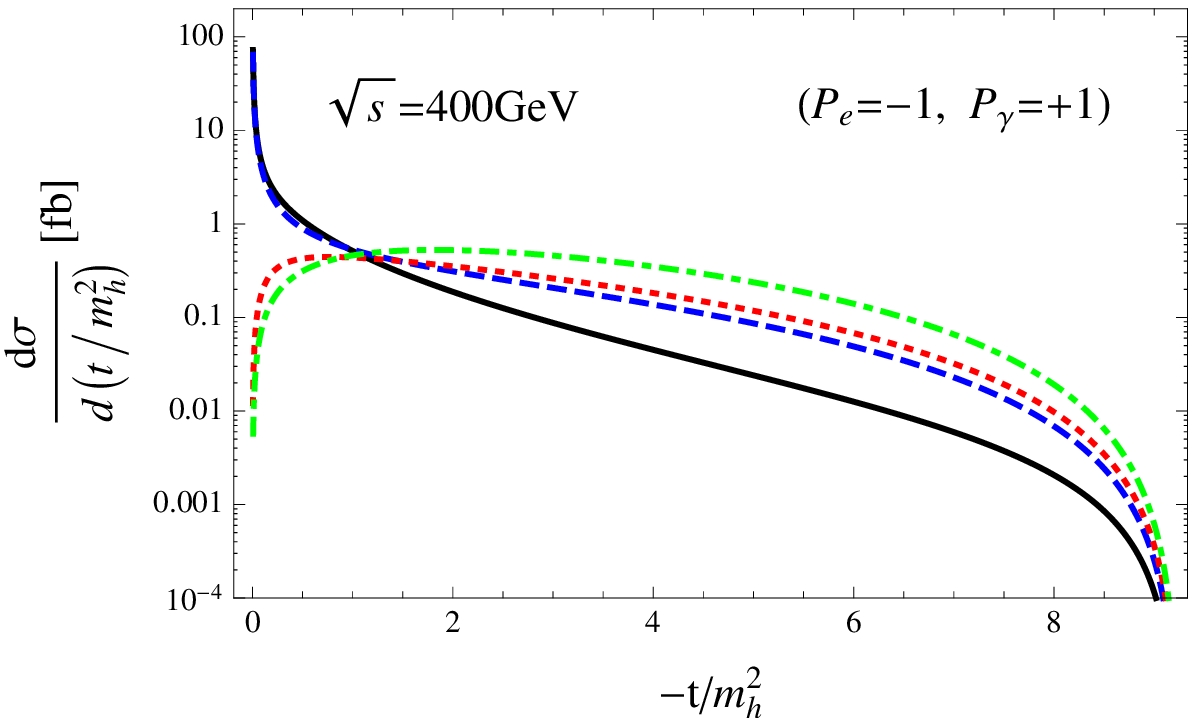}
 \end{center}
 \end{minipage}  &
\begin{minipage}{0.45\hsize}
 \begin{center}
  \includegraphics[width=80mm]{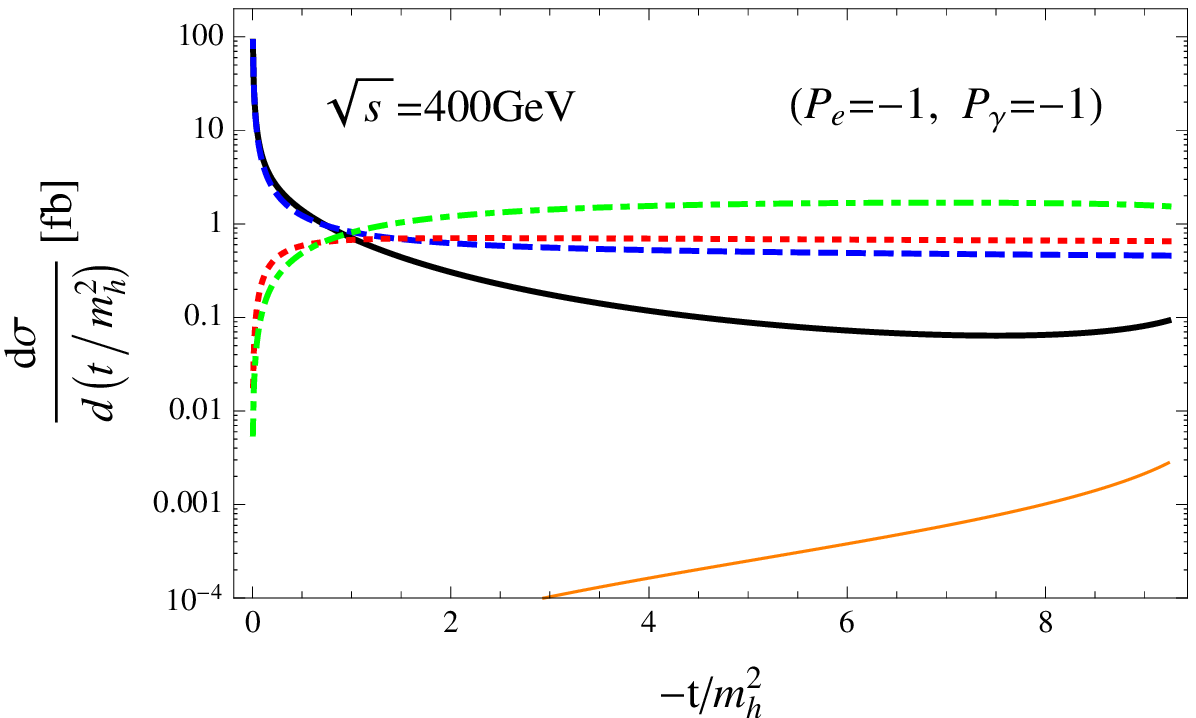}
 \end{center}
  \label{fig:three}
 \end{minipage}
 \end{tabular}
 \caption{The differential cross section for Higgs boson production $d\sigma_{(e\gamma\rightarrow eH)}/d(t/m_h^2)$  (black solid line) together with  $d\sigma_{(\gamma\gamma)}/d(t/m_h^2)$  (blue dashed line), $d\sigma_{(Z\gamma)}/d(t/m_h^2)$  (red dotted line), $d\sigma_{(W\nu_e)}/d(t/m_h^2)$  (green dot-dashed line) and $d\sigma_{(Ze)}/d(t/m_h^2)$  (orange thin solid line) as a function of
$-t/m_h^2$ with $\sqrt{s}=400$ GeV for four cases of  polarizations of the initial electron and photon beams, $(P_e=+1,P_\gamma=+1)$, $(P_e=+1,P_\gamma=-1)$, $(P_e=-1,P_\gamma=+1)$ and $(P_e=-1,P_\gamma=-1)$. In the plots of $(P_e=+1,P_\gamma=-1)$ and $(P_e=-1,P_\gamma=+1)$, $d\sigma_{(Ze)}/dt$ is too small and is out of the plot range.}
 \label{dsigmadt400}
\end{figure}

Now we analyze numerically the differential cross section $d\sigma_{(e\gamma\rightarrow eH)}(s,P_e,P_\gamma)/dt$ together with the other four differential cross sections given in Eqs.(\ref{dsigmadtgammagamma})-(\ref{dsigmadtZe}). 
We choose the mass parameters and the coupling constants as follows:
\bea
&& m_h=125~{\rm GeV}~,\quad m_t=173~{\rm GeV}~,\quad m_Z=91~{\rm GeV}~,\quad m_W=80~{\rm GeV}~,\label{Nume1}\\
&& \cos\theta_W=\frac{m_W}{m_Z}~,\qquad e^2=4\pi\alpha_{em}=\frac{4\pi}{128}~,\qquad 
g=\frac{e}{\sin\theta_W}~.\label{Nume2}
\eea
The electromagnetic constant $e^2$ is chosen to be the value at the scale of $m_Z$. 
We plot these differential cross sections 
as a function of $-t/m_h^2$ in Fig. \ref{dsigmadt200} and Fig. \ref{dsigmadt400} for the cases $\sqrt{s}= 200{\rm GeV}$ and $\sqrt{s}= 400{\rm GeV}$, respectively. The graphs are shown for each case of  polarizations of the electron and photon beams. First we find that the contribution from the ``$Ze$"  diagrams is very small for all cases compared with those from the other three. Actually, it is negligibly small when $P_eP_\gamma=-1$. In such cases, the terms with dominant $|S^{Ze}_{(k_{1})}(s,t,m_h^2,m_Z^2)|^2$ in Eq.(\ref{dsigmadtZe}) cancel out and $|S^{Ze}_{(k'_{1})}(s,t,m_h^2,m_Z^2)|^2$ vanishes as $u\rightarrow 0$  (or $t\rightarrow t_{\rm min}=m_h^2-s$). Also we see  that when $P_eP_\gamma=-1$ all the graphs indeed diminish as $u\rightarrow 0$. 

For the case of  polarizations $P_e=-1$ and $P_\gamma=\pm1$, 
a dominant contribution at smaller $|t|$, more specifically, up to  $-t/m_h^2= 1$, comes from the $\gamma^*\gamma$ fusion diagrams. This is due to the factor $(-1/t)$ in the expression (\ref{dsigmadtgammagamma}) for $d\sigma_{(\gamma\gamma)}/dt$, which arises as $(-t)\times(1/t^2)$ with $1/t$ coming from the photon propagator. For $1<-t/m_h^2< 1.5$, the contributions to the differential cross section from  $\gamma^*\gamma$ fusion, $Z^*\gamma$ fusion and ``$W\nu_e$"  diagrams become the same order, and at $-t/m_h^2>1.5$ (see Fig.\ref{dsigmadt400}), the contribution of  ``$W\nu_e$"  diagrams prevails over the other two, since ``$W\nu_e$"  diagrams do not have propagator factors such as $1/t$ and $1/(t-m_Z^2)$. For $P_e=-1$ and $P_\gamma=\pm1$, the interference between $A_{\gamma\gamma}$ and $A_{Z\gamma}$  works constructively, while the one between  $A_{\gamma\gamma}$  and $A_{W\nu_e}$ 
works destructively and its effect becomes large at $-t/m_h^2>1.5$. Thus the values of $d\sigma_{(e\gamma\rightarrow eH)}/dt$ become smaller than those of $d\sigma_{(\gamma\gamma)}/dt$, $d\sigma_{(Z\gamma)}/dt$ and $d\sigma_{(W\nu_e)}/dt$ (see Fig.\ref{dsigmadt400}).

For the electron polarization $P_e=+1$,  no contribution comes from ``$W\nu_e$"  diagrams. 
The interference between $A_{\gamma\gamma}$ and $A_{Z\gamma}$   for $P_e=+1$ and $P_\gamma=\pm1$ works destructively and its effect 
is large even for small $-t/m_h^2$. Therefore, $d\sigma_{(e\gamma\rightarrow eH)}/dt$ decreases rather rapidly as $-t/m_h^2$ increases.

Integrating the differential cross section given in Eq.(\ref{CrossSectionFormula}) over $t$, we obtain the Higgs boson production cross section
\bea
\sigma_{(e\gamma\rightarrow eH)}(s,P_e,P_\gamma)=\int_{\rm cut} dt\frac{d\sigma_{(e\gamma\rightarrow eH)}(s,P_e,P_\gamma)}{dt}~. \label{CrossSection}
\eea
It is known that the forward and backward directions in an $e^-\gamma$ collider are blind spots for the detection of scattered particles.  So we set kinematical cuts for the scattered electron in  $e^-\gamma$ collisions. We choose the allowed region of $\theta$ in the CM frame given in Eq.(\ref{CMangle}) as $10^\circ \le\theta\le170^\circ$, which leads to the integration range of $t$ in Eq.(\ref{CrossSection}) as $(-s+m_h^2-t_{\rm cut})\le t \le 
t_{\rm cut}$ with $t_{\rm cut}=-\frac{1}{2}(s-m_h^2)(1-\cos 10^\circ)$. We find that the imposition of  
kinematical cuts reduces the contribution of $\gamma^*\gamma$ fusion diagrams but has almost no effect 
on the contributions of the other $Z^*\gamma$ fusion, ``$W\nu_e$" and ``$Ze$"  diagrams.

\begin{figure}[htbp]
 \begin{tabular}{cc}
 \begin{minipage}{0.45\hsize}
  \begin{center}
   \includegraphics[width=80mm]{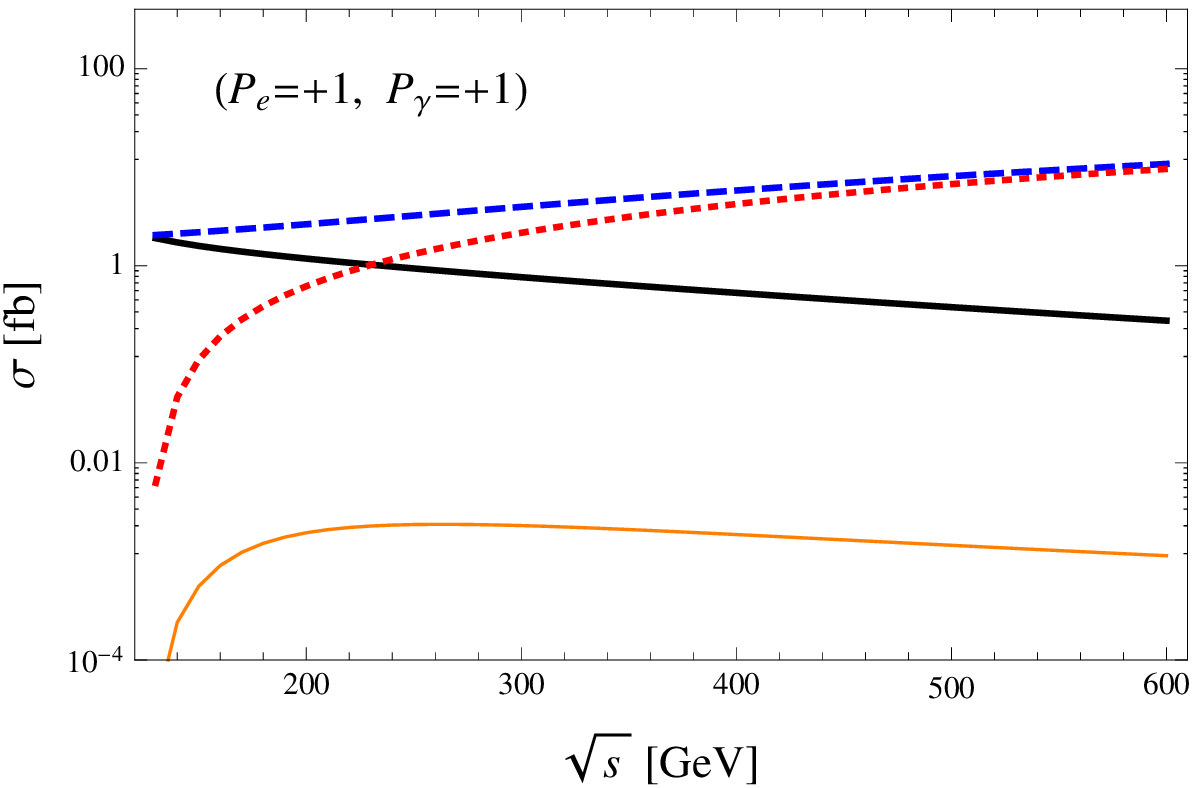}
  \end{center}
 \end{minipage} &
 \begin{minipage}{0.45\hsize}
 \begin{center}
  \includegraphics[width=80mm]{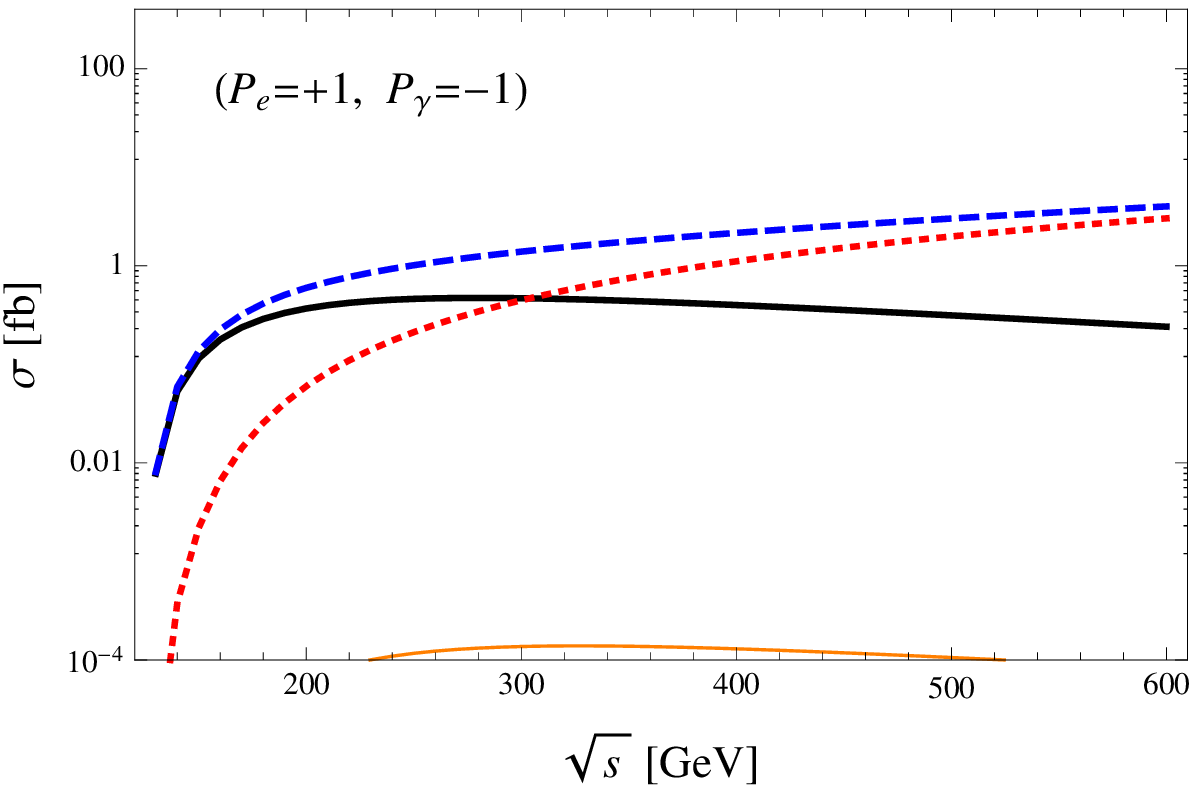}
 \end{center}
 \end{minipage}\\
 \begin{minipage}{0.45\hsize}
 \begin{center}
  \includegraphics[width=80mm]{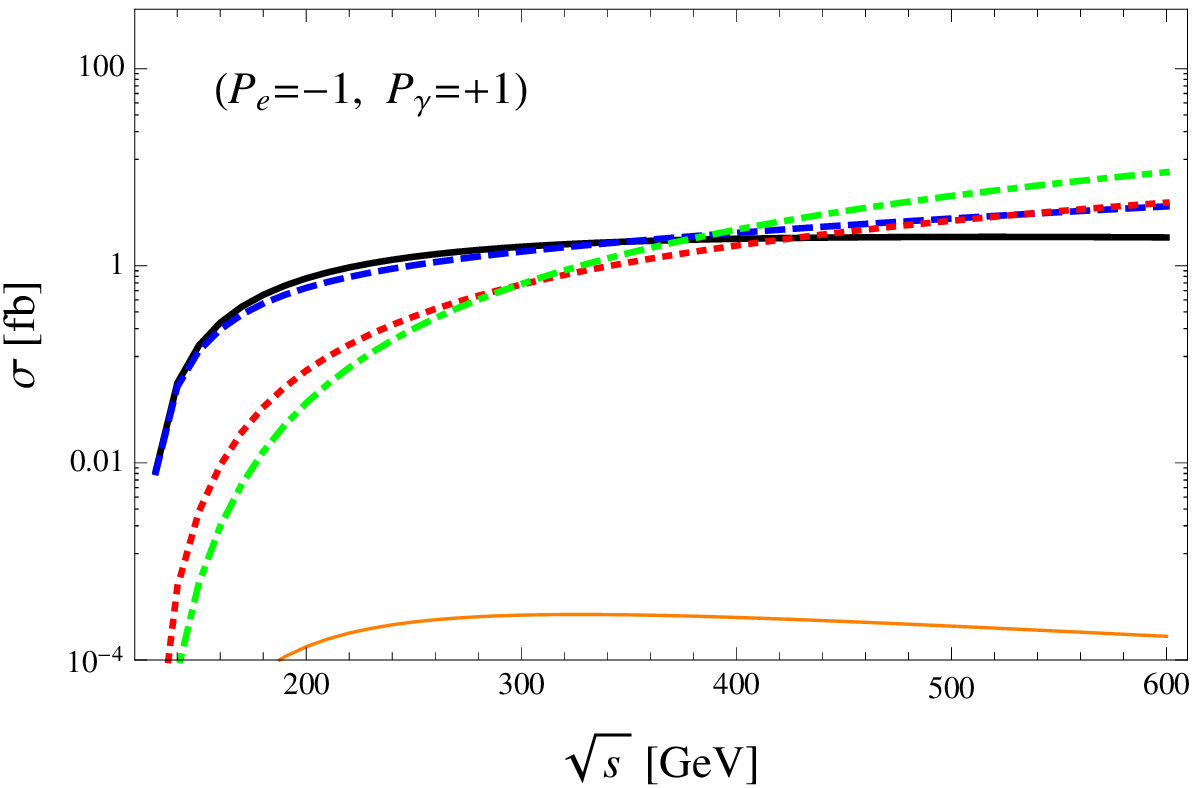}
 \end{center}
 \end{minipage}  &
\begin{minipage}{0.45\hsize}
 \begin{center}
  \includegraphics[width=80mm]{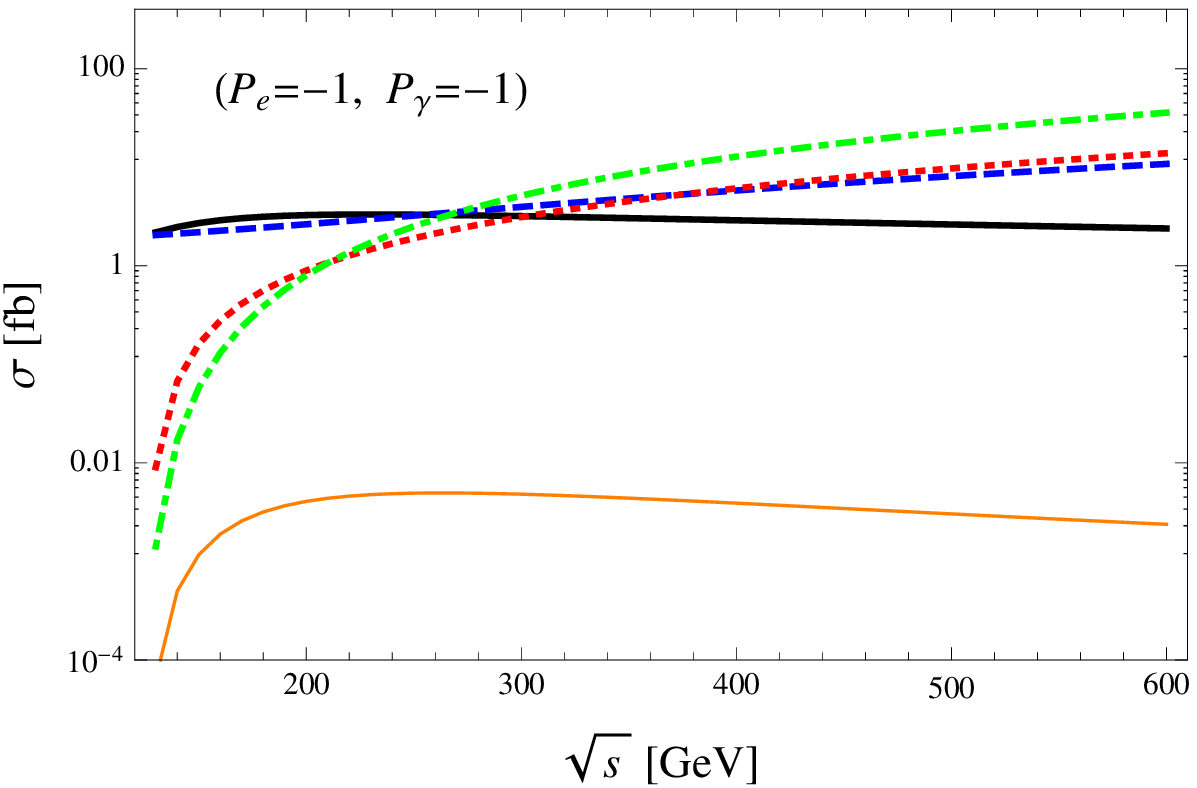}
 \end{center}
  \label{fig:three}
 \end{minipage}
 \end{tabular}
 \caption{Higgs boson production cross section $\sigma_{(e\gamma\rightarrow eH)}$ (black solid line), together with $\sigma_{(\gamma\gamma)}$ (blue dashed line), $\sigma_{(Z\gamma)}$ (red dotted line), $\sigma_{(W\nu_e)}$ (green dot-dashed line) and $\sigma_{(Ze)}$ (orange thin solid line ) as a function of $\sqrt{s}$ for four cases of  polarizations of the initial electron and photon beams, $(P_e=+1,P_\gamma=+1)$, $(P_e=+1,P_\gamma=-1)$, $(P_e=-1,P_\gamma=+1)$ and $(P_e=-1,P_\gamma=-1)$. The kinematical cut is chosen such that the allowed angle $\theta$ of the scattered electron in the CM frame is $10^\circ \le\theta\le170^\circ$.}
 \label{Crosssection}
\end{figure}

\begin{figure}[htbp]
  \begin{center}
   \includegraphics[width=80mm]{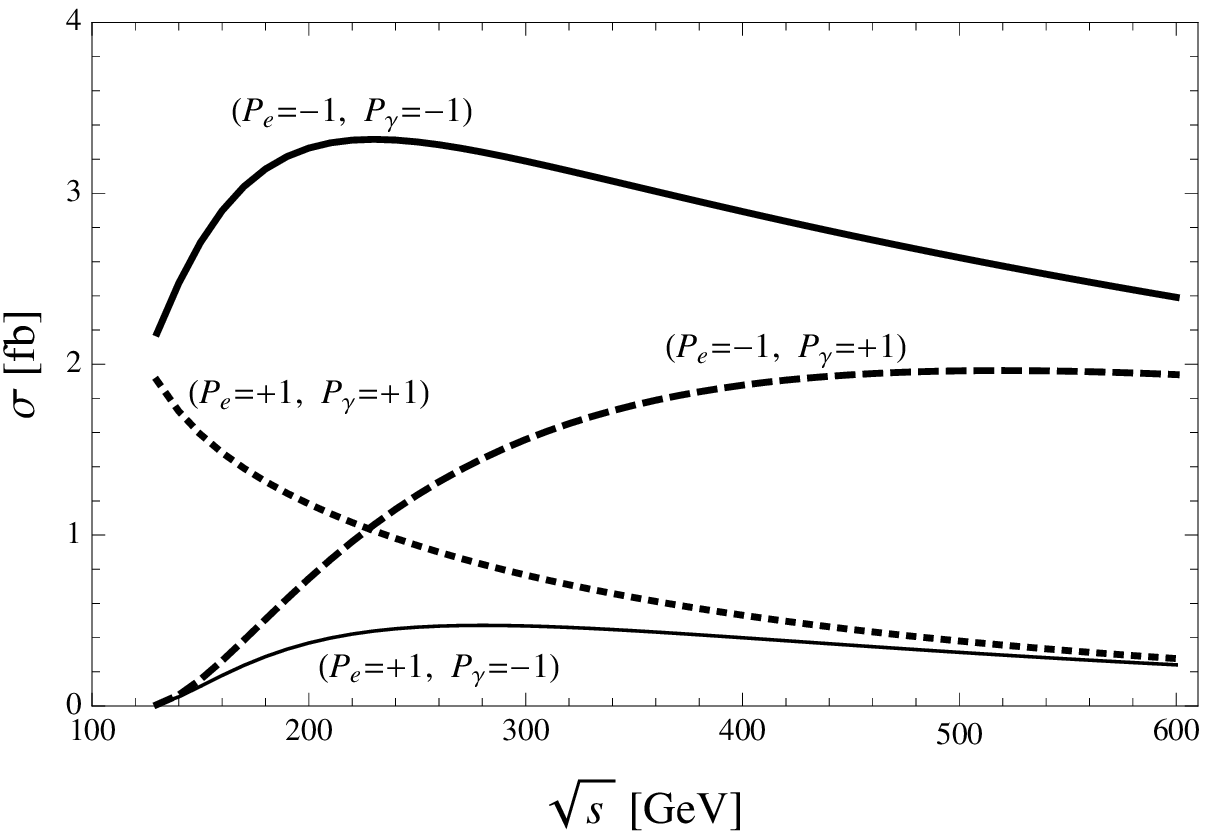}
  \end{center}
 \caption{Higgs boson production cross section $\sigma_{(e\gamma\rightarrow eH)}(s,P_e,P_\gamma)$ for the cases of $(P_e=-1,P_\gamma=-1)$ (thick solid line), $(P_e=-1,P_\gamma=+1)$ (dashed line), $(P_e=+1,P_\gamma=+1)$ (dotted  line ) and $(P_e=+1,P_\gamma=-1)$ (thin solid line). The kinematical cut is chosen such that the allowed angle $\theta$ of the scattered electron in the center-of-mass frame is $10^\circ \le\theta\le170^\circ$.}
 \label{CSsummary} 
\end{figure}

Similarly we define    $\sigma_{(\gamma\gamma)}(s,P_e,P_\gamma)$, $\sigma_{(Z\gamma)}(s,P_e,P_\gamma)$, $\sigma_{(W\nu_e)}(s,P_e,P_\gamma)$ and $\sigma_{(Ze)}(s,P_e,P_\gamma)$ by integrating the expressions given in Eqs.(\ref{dsigmadtgammagamma})-(\ref{dsigmadtZe}) over $t$. 
 We plot these cross sections in Fig.\ref{Crosssection} as a function of $\sqrt{s}$ ($\sqrt{s}\ge 130$ GeV) for each case of  polarizations of the electron and photon beams. The detailed behaviors of $\sigma_{(e\gamma\rightarrow eH)}(s,P_e,P_\gamma)$ in linear scale are summarized in Fig.\ref{CSsummary}. For the case $P_eP_\gamma =-1$,  
the Higgs boson production cross section $\sigma_{(e\gamma\rightarrow eH)}$ is very small at $\sqrt{s}=130$ GeV, since the integration range of $t$ is small and the differential cross section vanishes as 
$t\rightarrow t_{\rm min}$. The cross section $\sigma_{(e\gamma\rightarrow eH)}(s,P_e=-1,P_\gamma=+1)$
rises gradually up to about 2 fb, while $\sigma_{(e\gamma\rightarrow eH)}(s,P_e=+1,P_\gamma=-1)$ increases rather slowly up to 0.4 fb. This is due to the interference between 
$A_{\gamma\gamma}$ and $A_{Z\gamma}$, which acts constructively for ($P_e=-1,P_\gamma=+1$) but 
destructively for ($P_e=+1,P_\gamma=-1$). For the case $P_eP_\gamma =+1$,  
the cross section $\sigma_{(e\gamma\rightarrow eH)}$ is about 2 fb at $\sqrt{s}=130$ GeV. The cross section $\sigma_{(e\gamma\rightarrow eH)}(s,P_e=-1,P_\gamma=-1)$
rises  above 3 fb around $\sqrt{s}=200$ GeV and then gradually decreases as $\sqrt{s}$ increases.  This is due to the destructive interference both between $A_{W\nu_e}$ and $A_{\gamma\gamma}$ and  between $A_{W\nu_e}$ and $A_{Z\gamma}$ in the range of large $-t$.  
Again the destructive interference between $A_{\gamma\gamma}$ and $A_{Z\gamma}$ is responsible for the decrease of $\sigma_{(e\gamma\rightarrow eH)}(s,P_e=+1,P_\gamma=+1)$ as $\sqrt{s}$ increases.

\section{Analysis of Higgs boson production in $e^-\gamma$ collisions \label{Section4}} 
A high-intensity photon beam can be produced by laser light backward scattering off 
a high-energy electron beam, $e^-\gamma_{\rm Laser}\rightarrow e^-\gamma$, where the backward-scattered photon receives a major fraction of the  incoming electron energy~\cite{GinzburgTelnov}.
Its energy distribution depends on the polarizations of the initial electron 
($P_{e2}=\pm$) and laser photon ($P_{\rm Laser}=\pm$).
Assuming, for simplicity, that a low-energy laser photon (typically a few eV) has a head-on collision with an electron with high energy (typically 125-250 GeV), we  calculate the energy spectra
of the scattered photon for  different combinations  of  polarizations
of the initial beams and also of the scattered photon. In this situation, the laser photon is scattered 
backwards and gains a large portion of the electron energy. 
The energy spectrum of the 
scattered photon, which is the sum of two helicity states ($P_\gamma=\pm 1$), is given by~\cite{Sessler:1995er}: 
\bea
\frac{1}{\sigma_C}\frac{d\sigma_C}{dy}=\frac{\pi\alpha^2}{2E_{e2}E_{\rm{Laser}}\sigma_C}\left[\frac{1}{1-y}+1-y-4r(1-r)+P_{e2}P_{\rm Laser}
~rx(1-2r)(2-y)\right], \label{PhotonSpectrum}
\eea
where $E_{e2}$ and $E_{\rm{Laser}}$ are the energies of the initial electron and laser photon, respectively,  $\sigma_C$ is the total cross section of  Compton scattering and   $y=\frac{\omega}{E_{e2}}$ with $\omega$ being the energy of the scattered photon.  The  variable $r$ is defined as 
\bea
r=\frac{y}{x(1-y)},\qquad \qquad {\rm with}\quad x=\frac{4E_{e2}E_{\rm Laser}}{m_e^2}~,
\eea
where $m_e$ is the electron mass.
The maximum value of $y$ is given by
\bea
y_{\rm max}=\frac{x}{1+x}~.\label{Ymax}
\eea

\begin{figure}[htbp]
  \begin{center}
   \includegraphics[width=90mm]{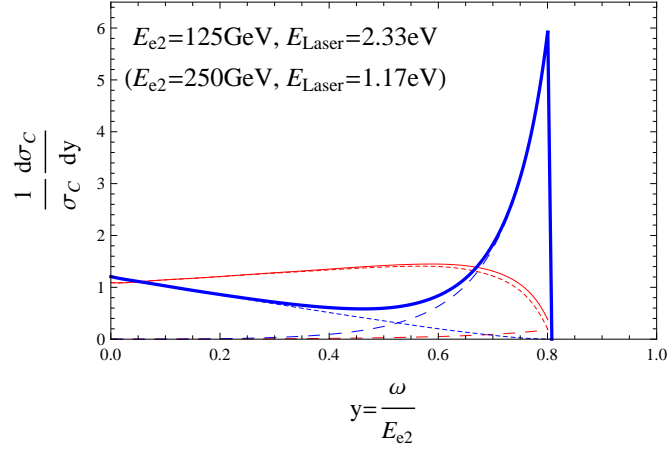}
  \end{center}
 \caption{The energy spectra of the scattered photons as a function of $y=\frac{\omega}{E_{e2}}$ 
for the cases of ($E_{\rm Laser}=2.33$eV, $E_{e2}=$125GeV) and ($E_{\rm Laser}=1.17$eV, $E_{e2}=$250GeV). 
For both cases, the energy spectra are described by the same graph.
The  blue and red  curves represent the spectra when the initial electron  and laser photon have the opposite polarizations ($P_{e2}P_{\rm Laser}=-1$) and the same 
polarizations ($P_{e2}P_{\rm Laser}=+1$), respectively.
The dashed and dotted curves show the helicity-flip  ($P_{\gamma}=-P_{\rm Laser}$) and  helicity-non-flip ($P_{\gamma}=P_{\rm Laser}$) components of the scattered photon, respectively, and the solid curves  represent the sum  of the two.
\label{polspectra}}
\end{figure}

The parameter $x$ should be less than 4.83, since the laser photons cannot be too energetic so that the scattered high-energy photons may not disappear by colliding with other laser photons to produce $e^+e^-$ pairs
\cite{Sessler:1995er}. 
We assume the use of laser photons with energy 2.33eV (corresponding to the YAG laser with wave-length 532nm) 
for the case of an electron beam with energy 125GeV and those with energy 1.17eV (the YAG laser with wave-length 1064nm) for an electron beam with energy 250GeV. For both cases we obtain $x=4.46$ and 
 the energy spectra of the scattered photons are described by the same graph shown in Fig.\ref{polspectra}. The solid blue and red  curves represent the spectra for the cases when the initial electron  and laser photon have
the opposite polarizations ($P_{e2}P_{\rm Laser}=-1$) and  the same 
polarizations ($P_{e2}P_{\rm Laser}=+1$), respectively. We see from Eq.(\ref{Ymax}) that  $82\%$ of the electron energy can be transferred to the scattered photon at the maximum. 

Actually we need the energy spectrum for each helicity state $P_\gamma$ of the scattered photon.  We use GRACE~\cite{GRACE} to calculate these two helicity components.
 The result is also shown  in Fig.{\ref{polspectra}}. 
 The dashed and dotted (blue and red) curves show the helicity-flip  ($P_{\gamma}=-P_{\rm Laser}$) and  helicity-non-flip ($P_{\gamma}=P_{\rm Laser}$) components of the scattered photon, respectively, and the solid curves  represent the sum  of the two, which are expressed by Eq.(\ref{PhotonSpectrum}).
It is noted that the spectrum with a peak at the kinematic endpoint, $y=y_{\rm max}$, is  obtained when $P_{e2}P_{\rm Laser}=-1$ (the thick solid blue curve in Fig.\ref{polspectra}). The highest-energy photons are produced by the helicity-flip process (the dashed curves) and their helicity $P_\gamma$ is the opposite of $P_{\rm Laser}$. Therefore, we are particularly  interested in the spectrum for the case $P_{e2}P_{\rm Laser}=-1$, where the helicity-flipped component dominates the large-$y$ region while helicity-conserved 
component occupies the small-$y$ region.

Suppose we have a highly polarized $e^-e^-$ collider machine. 
Converting one of the electron beams to photon beam by means of backward Compton scattering of a polarized laser beam, we obtain  an $e^-\gamma$ collider with high polarization.
Dividing the energy spectrum of the scattered photon, $\frac{1}{\sigma_C}\frac{d\sigma_C}{dy}$ given in Eq.(\ref{PhotonSpectrum}),  into two pieces $N(y,E_{e2},E_{\rm Laser},P_{e2},P_{\rm Laser},P_\gamma)$ depending on its helicity $P_\gamma$, the Higgs boson production cross section for $e^-\gamma\rightarrow e^-H$ in an $e^-e^-$ collider, whose beam energies are  $E_{e1}$ and $E_{e2}$ and polarizations are $P_{e1}$ and $P_{e2}$, is expressed as\footnote{Although,  in an actual laser backscattering,  the 
electron and laser beams intersect at a certain angle, here we assume a head-on collision of the two beams and use the  energy spectrum of the photon beam obtained by that assumption.},
\bea
&&\hspace{-2cm}\sigma_{e\gamma ~{\rm collision}}(s_{ee},E_{\rm Laser},P_{e1},P_{e2},P_{\rm Laser})\nn\\
&&=\sum_{P_{\gamma}}\int dy~N(y,E_{e2},E_{\rm Laser},P_{e2},P_{\rm Laser},P_\gamma)
\sigma_{(e\gamma\rightarrow eH)}({s,P_{e1},P_{\gamma}}),  \label{Convolution}
\eea
where $\sigma_{(e\gamma\rightarrow eH)}({s,P_{e1},P_{\gamma}})$ is given in Eq.(\ref{CrossSection}) with $P_e$ replaced 
by $P_{e1}$, and $s_{ee}$ is the CM energy squared of the two initial electron beams and is related to $s$ as $s=y s_{ee}$. The integration range of $y$ is given by $y_{\rm min} \le y\le y_{\rm max} $
with $y_{\rm min}=0.25 (0.0625)$ for the case $E_{e2}(=E_{e1})=125$GeV (250GeV).

\bigskip
 \begin{table}[htb]
 \scalebox{1.4}[1.5]
 {
  \begin{tabular}{|c|c|c||c|c|} \hline
   $ \sqrt{s_{ee}}~{\rm GeV} $& $P_{e1}$ &   $P_{\rm Laser}$ & 
   $\sigma_{\rm cut}~{\rm fb}$ & 
   $S/\sqrt{B}$ \\ \hline \hline
                   & 1 & -1 & 0.50 &6.17\\ \cline{2-5}
    $250$   & 1 & 1 & 0.36&4.48\\ \cline{2-5}
                   & -1 & -1 & 0.80 &4.51\\ \cline{2-5}
                   & -1 & 1 & 1.53 &8.68\\ \hline
                   & 1 & -1 & 0.11&2.93\\ \cline{2-5}
    $500$    & 1 & 1 & 0.19&1.31 \\ \cline{2-5}
                   & -1 & -1 & 1.22&10.6\\ \cline{2-5}
                   & -1 & 1 & 1.01 &6.8\\ \hline
        \end{tabular}
        }
 \caption{Higgs boson production cross section and significance~  in $e^-\gamma$ collision in an $e^-e^-$ collider for the cases 
(i) $E_{\rm Laser}=2.33$eV, $E_{e2}=$125GeV, $\sqrt{s_{ee}}=$250GeV and (ii) $E_{\rm Laser}=1.17$eV, $E_{e2}=$250GeV, $\sqrt{s_{ee}}=$500GeV, and for each combination of polarizations $P_{e1}$ and $P_{\rm Laser}$. $P_{e2}$ is chosen to be $-P_{\rm Laser}$.}
 \label{XSandSig}
\end{table}

A feasible channel to observe the SM Higgs boson with mass 125GeV is $b\bar{b}$ decay, since it has a 
large branching ratio. We analyze the cross section of the Higgs boson production through the $b\bar{b}$ decay channel, $e+\gamma\rightarrow e+H\rightarrow e+b+\bar{b}$. The energy spectrum of the photon beam is given by $N(y,E_{e2},E_{\rm Laser},P_{e2},P_{\rm Laser},P_\gamma)$ with $P_{e2}P_{\rm Laser}=-1$. We consider the two cases: (i) $E_{\rm Laser}=2.33$eV, $E_{e2}=$125GeV and (ii) $E_{\rm Laser}=1.17$eV, $E_{e2}=$250GeV. Both cases give the same spectrum. 
Note that we take 
the case $P_{e2}P_{\rm Laser}=-1$ so that the spectrum has a peak at the highest energy which corresponds to the blue solid curve in Fig.\ref{polspectra}.  The Monte Carlo method is used. 
A $b$-quark mass is chosen to be 4.3 GeV. The angle cuts of  the scattered electron and $b(\bar{b})$ quarks are chosen such that the allowed regions are $10^\circ \le\theta_{e^-}\le170^\circ$ and $10^\circ \le\theta_{b(\bar{b})}\le170^\circ$, respectively, and the energy cuts of these particles are set to be $3{\rm GeV}$.
The Monte Carlo statistical error is about $0.1\%$ when the sampling number is taken to be 200,000.

In Table \ref{XSandSig} we show the results of the Higgs boson production cross section $\sigma_{\rm cut}$ for  the cases $\sqrt{s_{ee}}=250{\rm GeV}$ and $\sqrt{s_{ee}}=500{\rm GeV}$ and for each combination of polarizations $P_{e1}$ and $P_{\rm Laser}$. In the case $\sqrt{s_{ee}}=250{\rm GeV}$ (and thus $E_{e2}$=125GeV), we obtain $y_{\rm min}=$0.25, $y_{\rm max}=$0.82, and  $\sqrt {s_{\rm max}}\equiv\sqrt{y_{\rm max}s_{ee}}$=226GeV. Hence the cross section $\sigma_{(e\gamma\rightarrow eH)}({s,P_{e1},P_{\gamma}})$ with $s\le s_{\rm max}$ is convolved with the photon energy spectrum in Eq.(\ref{Convolution}). The behaviors of $\sigma_{(e\gamma\rightarrow eH)}({s,P_{e},P_{\gamma}})$ for various polarizations $P_e$ and $P_\gamma$ with $\sqrt{s}$ below 226GeV which are shown in Fig.\ref{CSsummary} and the fact that  the helicity-flipped ($P_\gamma=-P_{\rm Laser}$) component (the dashed blue curve) has a peak at $y=y_{\rm max}$ and dominates the  spectrum region $0.5<y<y_{\rm max}$ (see Fig.\ref{polspectra}) lead to the expectation
\bea
\sigma_{\rm cut}(P_{e1}=-1,P_{\rm Laser}=1)&>&\sigma_{\rm cut}(P_{e1}=-1,P_{\rm Laser}=-1)\nn\\
&>&\sigma_{\rm cut}(P_{e1}=1,P_{\rm Laser}=-1)>\sigma_{\rm cut}(P_{e1}=1,P_{\rm Laser}=1)~,\nn
\eea 
for  $\sqrt{s_{ee}}=250{\rm GeV}$. The Monte Carlo results on $\sigma_{\rm cut}$ given in Table.\ref{XSandSig} confirm our expectation. 

On the other hand,  the Monte Carlo results for the case $\sqrt{s_{ee}}=500{\rm GeV}$ show 
$\sigma_{\rm cut}(P_{e1}=-1,P_{\rm Laser}=-1)>\sigma_{\rm cut}(P_{e1}=-1,P_{\rm Laser}=1)$ and $\sigma_{\rm cut}(P_{e1}=1,P_{\rm Laser}=1)>\sigma_{\rm cut}(P_{e1}=1,P_{\rm Laser}=-1)$.  The changes in the order of the sizes of the cross sections are explained as follows.  
For $\sqrt{s_{ee}}=500{\rm GeV}$, we obtain $y_{\rm min}=$0.0625, $y_{\rm max}=$0.82, and  $\sqrt {s_{\rm max}}$=452GeV. Fig.\ref{polspectra} tells us that the helicity-conserving component (the dotted blue curve) dominates 
in the small-$y$ region, i.e., $y_{\rm min}\le y< 0.5$. And we see from Fig.\ref{CSsummary} that both $\sigma_{(e\gamma\rightarrow eH)}(s,P_{e}=-1,P_{\gamma}=-1)$ and $\sigma_{(e\gamma\rightarrow eH)}(s,P_{e}=1,P_{\gamma}=1)$ increase as $\sqrt{s}$ decreases from $\sqrt {s_{\rm max}}$. These two factors give a concise account of 
the results on $\sigma_{\rm cut}$ for $\sqrt{s_{ee}}=500{\rm GeV}$ given in Table.\ref{XSandSig}.

So far we have considered the cases where both initial electrons and laser photons are 
fully polarized. Although we can prepare laser photons with full polarization, the electron 
beams are, in fact, partially polarized. Let us denote the polarization of the incident electrons as
\be
p\equiv p_R-p_L
\ee
where $p_R$ is the fraction with positive helicity and $p_L=1-p_R$ is the fraction with negative helicity. Now we reconsider the Higgs boson production cross section for $e^-\gamma\rightarrow e^-H$ in an $e^-e^-$ collider  when the two electron beams have polarizations $p_{e1}$ and $p_{e2}$. Then Eq.(\ref{Convolution}) is replaced by\footnote{Note that $P_{e1}$ and $P_{e2}$ take the fixed values $\pm 1$, while $p_{e1}$ and $p_{e2}$ are variables with the range $-1\le p_{e1}, p_{e2} \le 1$}
\bea
&&\hspace{-3cm}\sigma_{e\gamma ~{\rm collision}}(s_{ee},E_{\rm Laser},p_{e1},p_{e2},P_{\rm Laser})\nn\\
&&\hspace{-2cm}=\sum_{P_{\gamma}}\int dy~\Bigl\{\frac{1+p_{e2}}{2}N(y,E_{e2},E_{\rm Laser},P_{e2}=+1,P_{\rm Laser},P_\gamma)\nn\\
&&\qquad +\frac{1-p_{e2}}{2}N(y,E_{e2},E_{\rm Laser},P_{e2}=-1,P_{\rm Laser},P_\gamma)
\Bigr\}\nn\\
&&\times \Bigl\{\frac{1+p_{e1}}{2}
\sigma_{(e\gamma\rightarrow eH)}({s,P_{e1}=+1,P_{\gamma}})+
\frac{1-p_{e1}}{2}
\sigma_{(e\gamma\rightarrow eH)}({s,P_{e1}=-1,P_{\gamma}})
\Bigr\}~.
\eea
For example, setting $p_{e1}=p_{e2}=-0.8$~\cite{Kurihara} and $P_{\rm Laser}=+1$, we obtain 
$\sigma_{\rm cut}=1.28$fb and 0.92 fb, respectively, for the case $\sqrt{s_{ee}}=250{\rm GeV}$ and 
$\sqrt{s_{ee}}=500{\rm GeV}$. Particularly, here
we have chosen a negative polarization for $p_{e1}$. This is in order to increase the contribution from the ``$W\nu_e$" one-loop diagrams.

\begin{figure}[htbp]
  \begin{center}
   \includegraphics[width=80mm]{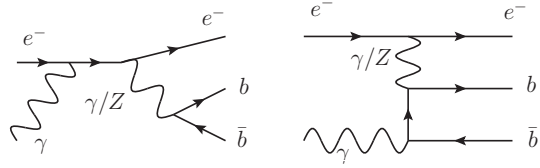}
  \end{center}
 \caption{Examples of background processes for $e+\gamma \rightarrow e+b+\bar{b}$.}
 \label{Background} 
\end{figure}

We also analyze the significance $S/\sqrt{B}$ of the Higgs boson production in $e^-\gamma$ collisions. The $b\bar{b}$ decay 
channel of the Higgs boson in $e^-\gamma$ collisions has a substantial background. Two examples of the 
background processes are shown in Fig.\ref{Background}. 

\noin
In particular, a huge background appears at the $Z$-boson pole. However, it is expected that when we measure the invariant mass $m_{b\bar{b}}$ of $b$ and $\bar{b}$ quarks, the background will be small in the region $m_{b\bar{b}}>$120GeV compared with the signals of the Higgs boson production. We use GRACE to write down all the tree Feynman diagrams for $e+\gamma \rightarrow e+b+\bar{b}$ and to evaluate their contributions to the background cross section. We assume that the integrated luminosity is 250${\rm fb}^{-1}$.
The significance is calculated by taking samples in the region 
$120{\rm GeV}\le m_{b\bar{b}}\le130{\rm GeV}$ at the parton level. The results are given 
in Table \ref{XSandSig}. Large values of significance are obtained for the cases of 
$(P_{e1},P_{{\rm Laser}})=(1,-1)$ and $(-1,1)$ with $\sqrt{s_{ee}}=250{\rm GeV}$ 
and $(P_{e1},P_{{\rm Laser}})=(-1,\mp 1)$ with $\sqrt{s_{ee}}=500{\rm GeV}$.

An additional background to the reaction $e+\gamma \rightarrow e+H \rightarrow e+b+\bar{b}$ is the resolved photon process 
$e+\gamma(g) \rightarrow e+b+\bar{b}$, where the gluon content in a photon interacts with $\gamma$ (and also $Z$) to produce a $b\bar{b}$ pair. This background process was considered in Refs.~\cite{Resolved,Gabrielli}.
We estimate this background as follows. The gluon content in a photon has not been 
reliably measured until now and parametrizations of the gluon distribution have been proposed only for  the case of an unpolarized photon in the literature~\cite{GinGamma,GRV}.
Therefore, we study  both the resolved photon process $e+\gamma(g) \rightarrow e+b+\bar{b}$ 
and the direct photon process $e+\gamma \rightarrow e+b+\bar{b}$ in an $e^-e^-$ collider for the case when  initial electron beams and laser photons are unpolarized. We use the gluon distribution function $f_{G/\gamma}(z,Q^2)$ in an unpolarized photon  given by the parametrization in Ref.\cite{GRV}, where $Q^2$ is the scale at which the structure of the photon is being probed and  $z$ is the fraction of the photon energy carried by the gluon.
The cross section for  $e+\gamma(g) \rightarrow e+b+\bar{b}$  is expressed as
\bea
\int^{y_{\rm max}}_0 dy N_{\rm unpol}(y,E_{e2},E_{laser})\int^1_0 dzf_{G/\gamma}(z,Q^2)~ \sigma(eg\rightarrow eb \bar b)(z y s_{ee})
\eea
where $N_{\rm unpol}(y,E_{e2},E_{laser})=\frac{1}{\sigma_C}\frac{d\sigma_C}{dy}|_{\rm unpol}$~which is obtained from Eq.(\ref{PhotonSpectrum}) by setting $P_{e2}P_{\rm Laser}=0$. 
Again we use GRACE to write down all the tree Feynman diagrams for $e+g \rightarrow e+b+\bar{b}$ and to evaluate the cross section $\sigma(eg\rightarrow eb \bar b)$. 
We choose $Q^2$ as $-(k_1-k_1')^2=-t$ for $f_{G/\gamma}(z,Q^2)$ and also for the running strong coupling constant $\alpha_s(Q^2)$.
Since the gluon content in a photon accumulates in a small-$z$ region, the invariant mass $m_{b\bar{b}}$ 
distribution of the resolved photon background cross section is expected to become smaller as  $m_{b\bar{b}}$ gets large.
It is also noted that in the direct photon process, the effect of the $Z$-boson pole (see the left diagram in Fig.\ref{Background}) on the background cross section remains to some extent as a tail in the region $m_{b\bar{b}}=125$ GeV, but that in the resolved 
photon process, there is no such  $Z$-boson pole effect.  
We obtain 0.06 for the background ratio of $d\sigma_{\rm cut}/d m_{b\bar{b}}$ between the resolved photon and direct photon processes at $m_{b\bar{b}}=125$ GeV  in the case $\sqrt{s_{ee}}=500$ GeV. When $\sqrt{s_{ee}}=250$GeV, the ratio becomes negligibly small.
Hence we find that the contribution of the resolved photon process to the background is very small compared to that of the direct photon 
process when we observe a pair of $b\bar{b}$ at the invariant mass around $m_{b\bar{b}}=125$GeV.
Although the analysis so far was on the background contributions for the case 
of the unpolarized beams, we expect  that the same trend still remains when we use
the polarized electron and photon beams.

Due to the electric charge factors, $c\bar c$ pairs have larger production cross sections than $b\bar b$ pairs.
If the $c\bar c$ pairs are misidentified as $b\bar b$, they turn out to be a further background for the reaction 
$e+\gamma \rightarrow e+H \rightarrow e+b+\bar{b}$. This is a reducible background and can be controllable if we have a detector with good efficiency for $b$ identification and $c$ rejection. We compute the rate for $e+\gamma \rightarrow e+c+\bar{c}$ 
in an $e^-e^-$ collider for the case of the unpolarized electron and photon beams. We find that $d\sigma_{\rm cut}(e\gamma \rightarrow ec\bar{c})/d m_{c\bar{c}}$ is  larger than 
$d\sigma_{\rm cut}(e\gamma \rightarrow eb\bar{b})/d m_{b\bar{b}}$ at $m_{c\bar{c}}=m_{b\bar{b}}=125$GeV by a factor of 3.2 (3.1) when $\sqrt{s_{ee}}=250$ GeV ($\sqrt{s_{ee}}=500$ GeV).
The effect of electric charge difference,  $Q_c=\frac{2}{3}$ and $Q_b=-\frac{1}{3}$, does not appear as 
significant as we have expected.  The reason is that, as mentioned above,  at the invariant mass of 125 GeV, the contribution from the diagrams with a $Z$-boson propagator also becomes  important. 
Thus  this reducible background $e\gamma \rightarrow ec\bar{c}$ will be brought under control once we prepare for a  detector with a good $b$-tagging  efficiency.

After all, we conclude 
that the Higgs boson will be clearly observed in $e^-\gamma$ collision experiments

\begin{figure}[htbp]
 \begin{tabular}{c}
 \begin{minipage}{0.5\hsize}
  \begin{center}
   \includegraphics[width=75mm]{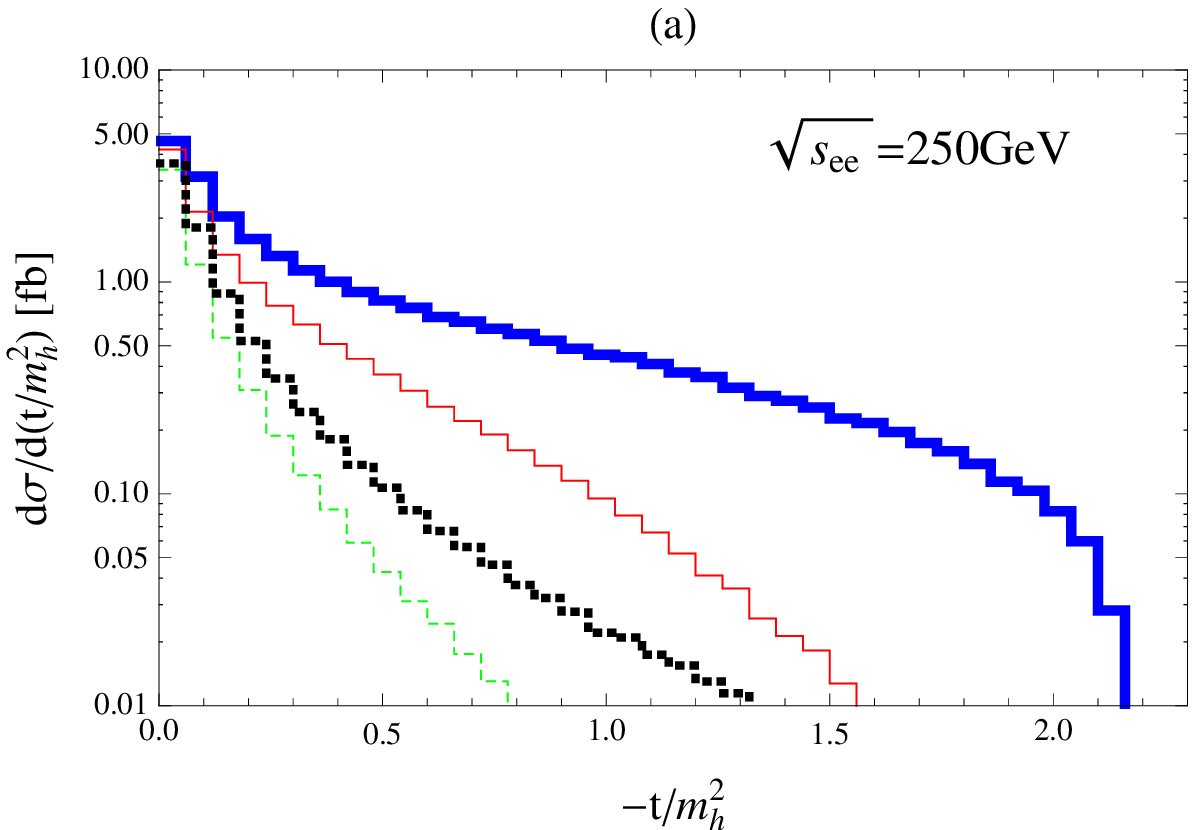}
  \end{center}
 \end{minipage}
 \begin{minipage}{0.5\hsize}
  \begin{center}
   \includegraphics[width=75mm]{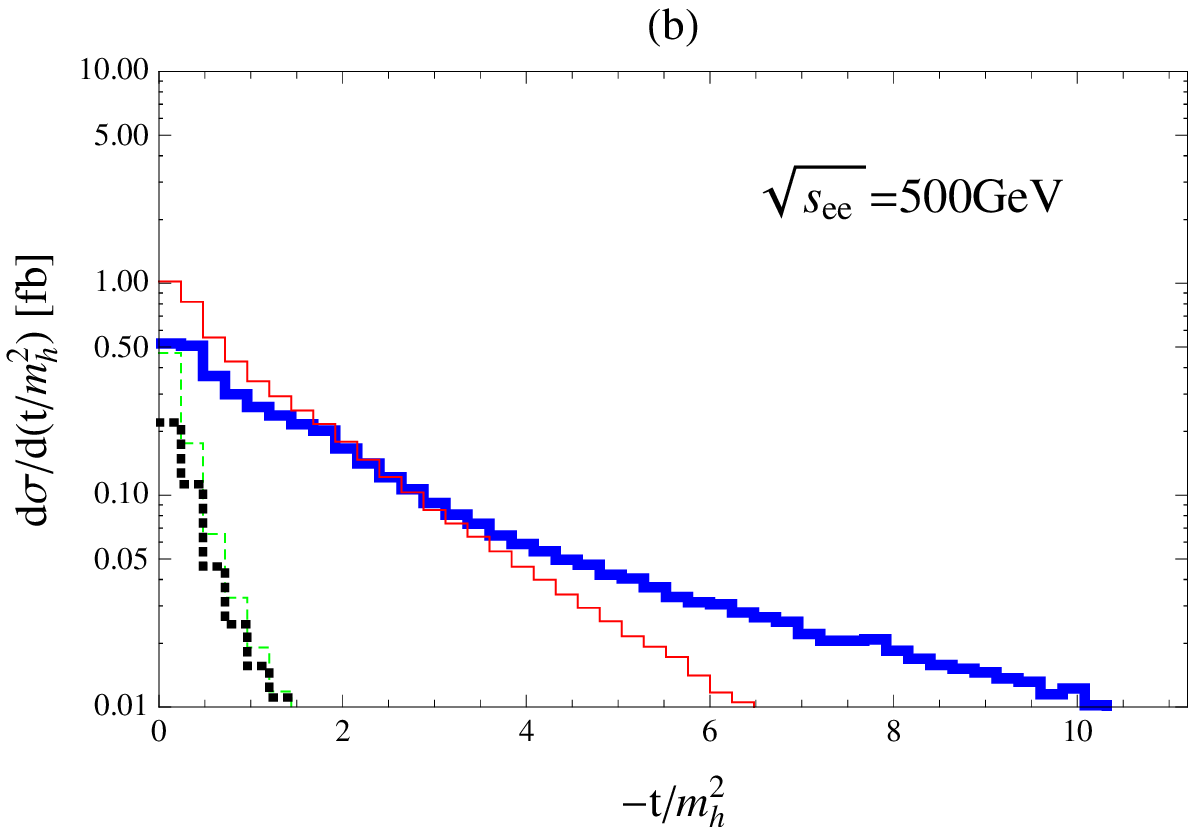}
  \end{center}
 \end{minipage}
 \end{tabular}
\caption{ Monte Carlo results for the differential cross section $d\sigma_{\rm cut}/d(t/m_h^2)$ 
 for the process $e\gamma\rightarrow eH\rightarrow e(b\bar{b})$  in $e^-\gamma$ collision in an $e^-e^-$ collider  for  the cases (a) $\sqrt{s_{ee}}=250{\rm GeV}$ and (b) $\sqrt{s_{ee}}=500{\rm GeV}$. 
The blue thick, red thin, black thick dashed and green thin dashed  lines represent the results for the cases where $(P_{e1},P_{{\rm Laser}})$ are $(-1,1)$, $(-1,-1)$, $(1,-1)$ and $(1,1)$, respectively, and $P_{e2}=-P_{\rm Laser}$. }
\label{difXS}
\end{figure}

Finally we show the results of our Monte Carlo analysis  on the differential cross section $d\sigma_{\rm cut}/d(t/m_h^2)$ for the process $e\gamma\rightarrow eH\rightarrow e(b\bar{b})$  in $e^-\gamma$ collision in an $e^-e^-$ collider.  In Fig.\ref{difXS} we plot  $d\sigma_{\rm cut}/d(t/m_h^2)$ as a function of $-t/m_h^2$ for  the cases (a) $\sqrt{s_{ee}}=250{\rm GeV}$ and (b) $\sqrt{s_{ee}}=500{\rm GeV}$, and for each combination  of polarizations $P_{e1}$ and $P_{\rm Laser}$. The kinematical cuts are the same as before. The behaviors of the four differential cross sections in Fig.\ref{difXS}(a) 
are consistent with  the observation that, for $\sqrt{s_{ee}}=250{\rm GeV}$, the helicity-flipped ($P_\gamma=-P_{\rm Laser}$) component dominates the  photon spectrum and with the results on  $d\sigma_{(e\gamma\rightarrow eH)}/d(t/m_h^2)$  (black solid lines) for $\sqrt{s}=$200GeV in Fig.\ref{dsigmadt200}. 
Also these four differential cross sections are in conformity with the numerical values of 
$\sigma_{\rm cut}$(fb) for  $\sqrt{s_{ee}}=250{\rm GeV}$ in Table \ref{XSandSig}. In the case 
$\sqrt{s_{ee}}=500{\rm GeV}$, the lower region of y, i.e., $0.0625<y<0.5$ where the helicity-conserving component  dominates  the photon spectrum, participates  in the convolution integral in Eq.(\ref{Convolution}) as well. Together with the results on $d\sigma_{(e\gamma\rightarrow eH)}/d(t/m_h^2)$ 
 (black solid lines) for $\sqrt{s}=$400GeV in Fig.\ref{dsigmadt400}, this explains the behaviors of the four differential cross sections in Fig.\ref{difXS}(b).  The crossover of the blue thick $(P_{e1}=-1,P_{\rm Laser}=1)$ and red thin $(P_{e1}=-1,P_{\rm Laser}=-1)$
lines near $-t/m_h^2=3$ gives an account of the change in the order of the size for $\sigma_{\rm cut}$ in Table \ref{XSandSig}.
We see a sharp drop of $d\sigma_{\rm cut}/d(t/m_h^2)$ for the case $P_{e1}=1$ (see the black thick dashed and green thin dashed
lines in Fig.\ref{difXS} (a) and (b)). This is due to the fact that, for the case $P_{e1}=1$, no contribution comes from ``$W\nu_e$"  diagrams and the interference between $\gamma^*\gamma$ and $Z^*\gamma$ fusion diagrams works destructively.

In Ref.\cite{KWSUPL}, we have pointed out the transition 
form factor of the Higgs boson via $\gamma^*\gamma$ fusion and its feasibility of observation in $e^-\gamma$ collision experiments.
In Table \ref{XSandSig} we see rather large significances for both $\sqrt{s_{ee}}=250{\rm GeV}$ and $\sqrt{s_{ee}}=500{\rm GeV}$ in the case of $P_{e1}=-1$. See also the blue and red plots in Fig.\ref{difXS} (a) and (b). As regards the differential cross section for the Higgs boson production, we find that 
the contribution of $\gamma^*\gamma$ fusion diagrams is dominant  up to $-t/m_h^2=1$ for the case of $P_{e1}=-1$. Hence we conclude that when the left-handed  
electron beam is used, the transition form factor of the Higgs boson is measurable 
 and  extracted 
from the differential cross section for the Higgs boson production up to $-t/m_h^2=1$.

\section{Summary \label{FinalSection}}
We  have investigated the  SM Higgs boson production in  $e^-\gamma$ collisions. The electroweak one-loop 
contributions to the scattering amplitude for $e^-\gamma\rightarrow e^-H$ were calculated and they were 
expressed in analytical form. Since large polarizations for the initial beams can be obtained in linear colliders, we analyzed 
both the differential cross section $d\sigma_{(e\gamma\rightarrow eH)}(s,P_e,P_\gamma)/dt$ and  the    cross section $\sigma_{(e\gamma\rightarrow eH)}(s,P_e,P_\gamma)$ for each combination of  polarizations of the  electron and photon beams.  We have found the following.  (i) Both the differential cross section and cross section are significantly dependent on the polarizations of the  electron and photon beams; (ii) The interferences between $\gamma^*\gamma$ and $Z^*\gamma$ fusion diagrams and between $\gamma^*\gamma$ fusion and ``$W\nu_e$"  diagrams, which work destructively or constructively depending on the polarizations of the initial beams, are important factors affecting the behaviors of both the differential cross section and cross section; (iii) For  $P_e=-1$, the contribution to $d\sigma_{(e\gamma\rightarrow eH)}/dt$ from  $\gamma^*\gamma$ fusion diagrams is dominant for  $-t/m_h^2\le 1$. Thus the transition form factor of the Higgs boson is measurable and extracted from the differential cross section for the Higgs boson production up to $-t/m_h^2=1$, when the left-handed  
electron beam is used; (iv) The ``$W\nu_e$" diagrams do not contribute to the reaction $e^-\gamma\rightarrow e^-H$ for $P_e=+1$. But they take part in the reaction  for $P_e=-1$, together with the $\gamma^*\gamma$ fusion and $Z^*\gamma$ fusion diagrams.  (v) the contribution from 
``$Ze$"  diagrams is extremely small and can be negligible. 

We analyzed the cross section of the Higgs boson production through the $b\bar{b}$ decay channel, $e+\gamma\rightarrow e+H\rightarrow e+b+\bar{b}$, in an $e^-\gamma$ collision in an $e^-e^-$ collider. 
A high-energy photon beam was assumed to be produced by laser light backward scattering off one of the high-energy electron beams of the 
$e^-e^-$ collider. We obtained large values of the significance $\sqrt{S}/B$ for the Higgs boson production for 
 both $\sqrt{s_{ee}}=250{\rm GeV}$ and  $\sqrt{s_{ee}}=500{\rm GeV}$. We therefore conclude that the Higgs boson will be clearly observed in $e^-\gamma$ collision experiments.

As a final comment,  we point out that in an $e^-\gamma$ collider, the photon structure functions can be measured by the single electron-tagging experiments. Analyses of photon structure functions have 
been intensively performed by using perturbative QCD  \cite{Uematsu}.


\begin{acknowledgments}
We thank the organizers of QFTHEP'2013 and RADCOR 2013 for the hospitality and the pleasant atmosphere during the workshops. We also thank Edward Boos, Mikhail N.~Dubinin and Ilya F.~Ginzburg for useful discussions.
\end{acknowledgments}



\newpage

\appendix

\section{Feynman Rules\label{FeynmanRules}}

\noin
The $W$- and $Z$-boson propagators in unitary gauge are, respectively, given by
\be
\frac{-i}{k^2-m_W^2}\Bigl(g_{\mu\nu}-\frac{k_\mu k_\nu}{m_W^2}\Bigr)~,\qquad 
\frac{-i}{k^2-m_Z^2}\Bigl(g_{\mu\nu}-\frac{k_\mu k_\nu}{m_Z^2}\Bigr)~.
\ee

\noin
Feynman rules for the tree-point and four-point vertices,
\bea
e\cdot e\cdot \gamma~  {\rm vertex}&:&\quad i(-e) \gamma_\mu~\\
t\cdot t\cdot \gamma~  {\rm vertex}&:&\quad i(Q_t e) \gamma_\mu\\
e\cdot \nu\cdot W~  {\rm vertex}&:&\quad i\frac{g}{2\sqrt{2}} \gamma_\mu(1-\gamma_5)\\
e\cdot e\cdot Z~  {\rm vertex}&:&\quad i\frac{g}{4\cos\theta_W}\gamma_\mu(f_{Ze}+\gamma_5)\qquad {\rm with}\qquad f_{Ze}=-1+4\sin^2\theta_W\\
t\cdot t\cdot Z~  {\rm vertex}&:&\quad i~\frac{g}{4\cos\theta_W}\gamma_\mu~[f_{Zt}-\gamma_5]\qquad {\rm with}\qquad f_{Zt}=1-\frac{8}{3}\sin^2\theta_W \label{ttZcoupling}\\
&&\nn\\
{\rm Higgs}\cdot t\cdot t~  {\rm vertex}&:&\quad  -i~\frac{g m_t}{2m_W}\\
{\rm Higgs}\cdot W\cdot W~  {\rm vertex}&:&\quad igm_W g_{\mu\nu}\\
{\rm Higgs}\cdot Z\cdot Z~  {\rm vertex}&:&\quad i\frac{gm_Z}{\cos\theta_W} g_{\mu\nu}\\
&&\nn\\
A_\mu(k_1)\cdot W^+_\nu(k_2)\cdot W^-_\lambda(k_3)~{\rm vertex}&:&\quad -ie\Bigl[(k_1-k_2)_\lambda g_{\mu\nu}+(k_2-k_3)_\mu g_{\nu\lambda}+(k_3-k_1)_\nu g_{\lambda\mu}\Bigr]\label{AWWvertex}\\
Z_\mu(k_1)\cdot W^+_\nu(k_2)\cdot W^-_\lambda(k_3)~{\rm vertex}&:&\quad -ig\cos\theta_W\Bigl[(k_1-k_2)_\lambda g_{\mu\nu}+(k_2-k_3)_\mu g_{\nu\lambda}+(k_3-k_1)_\nu g_{\lambda\mu}\Bigr]\label{ZWWvertex}\\
&&\nn\\
A_\mu\cdot A_\nu\cdot W^+_\alpha\cdot W^-_\beta ~{\rm vertex}&:& \quad
 -ie^2\Bigl[2 g_{\mu\nu}g_{\alpha\beta}-g_{\mu\alpha}g_{\nu\beta}- g_{\mu\beta}g_{\nu\alpha}\Bigr]\\
A_\mu\cdot Z_\nu\cdot W^+_\alpha\cdot W^-_\beta ~{\rm vertex}&:& \quad
-ieg\cos\theta_W\Bigl[2 g_{\mu\nu}g_{\alpha\beta}-g_{\mu\alpha}g_{\nu\beta}- g_{\mu\beta}g_{\nu\alpha}\Bigr]~.
\eea
where, in Eqs.(\ref{AWWvertex}) and (\ref{ZWWvertex}), momenta are all inward. 

\section{Scalar  One-loop Integrals\label{OLI}}

The scalar one-loop integrals which appeared in Section \ref{Section2} are the two-, three- and four-point 
integrals which are defined as
\bea
B_0(p^2; m_1^2, m_2^2)&\equiv&\frac{(2\pi\mu)^{4-n}}{i\pi^2}\int\frac{d^n k}{\Bigl[k^2-m_1^2\Bigr]\Bigl[(k+p)^2-m_2^2\Bigr]}\\
&&\hspace{-5cm}C_0(p_1^2, p_2^2, p_3^2; m_1^2, m_2^2, m_3^2)\equiv\frac{(2\pi\mu)^{4-n}}{i\pi^2}\int\frac{d^n k}{\Bigl[k^2-m_1^2\Bigr]\Bigl[(k+p_1)^2-m_2^2\Bigr]\Bigl[(k+p_1+p_2)^2-m_3^2\Bigr]}\\
&&\hspace{-5cm}D_0(p_1^2, p_2^2, p_3^2,p_4^2; s_{12},s_{23};m_1^2, m_2^2, m_3^2,m_4^2)\nn\\
&&\hspace{-3.8cm}\equiv\frac{(2\pi\mu)^{4-n}}{i\pi^2}\int\frac{d^n k}{\Bigl[k^2-m_1^2\Bigr]\Bigl[(k+p_1)^2-m_2^2\Bigr]\Bigl[(k+p_1+p_2)^2-m_3^2\Bigr]\Bigl[(k-p_4)^2-m_4^2\Bigr]}~,
\eea
where $n=4-2\epsilon$ and $\mu$ is the mass scale of dimensional regularization. Note that  $p_1+p_2+p_3=0$ for the three-point function $C_0$ and 
$p_1+p_2+p_3+p_4=0$ for the four-point function $D_0$. We evaluate these integrals for the case of the parameters $s,t,u,m_h^2, m_W^2$ and $m^2_Z$ which satisfy the following kinematical constraints:
\bea
&&s> m_h^2~,\qquad -(s-m_h^2)\le t\le 0~, \qquad -(s-m_h^2)\le u\le 0\nn\\
&&m_h^2<4m_W^2~,\qquad m_h^2<4m_Z^2~.
\eea
Therefore, analytic continuation is necessary for the variable $s$ from $s<0$ to $s>m_h^2$.\\

The integrals are expressed in terms of the following ratios:
\bea
t_T&\equiv& \frac{t}{m_t^2}~,\quad h_T\equiv\frac{m_h^2}{m_t^2}\\
s_W&\equiv& \frac{s}{m_W^2}~,\quad t_W\equiv \frac{t}{m_W^2}~,\quad u_W\equiv \frac{u}{m_W^2}~,\quad h_W\equiv\frac{m_h^2}{m_W^2}\\
s_Z&\equiv& \frac{s}{m_Z^2}~,\quad t_Z\equiv \frac{t}{m_Z^2}~,\quad u_Z\equiv \frac{u}{m_Z^2}~,\quad h_Z\equiv\frac{m_h^2}{m_Z^2}~.
\eea

\subsection{Two-point integrals}
The ultraviolet divergences appear in the scalar two-points integrals $B_0$'s 
and they are expressed by the  $\frac{1}{\epsilon}$ terms in dimensional regularization. 
But when we take difference between two $B_0$'s, the result becomes finite. Specifically we obtain
\bea
B_0(m_h^2;m_t^2,m_t^2)-B_0(t;m_t^2,m_t^2)&=&-2\sqrt{\frac{4}{h_T}-1}~\sin^{-1}\Bigl(\sqrt{\frac{h_T}{4}}\Bigr)\nn\\
&&+\sqrt{1-\frac{4}{t_T}}~\log\Bigl(\frac{\sqrt{4-t_T}+\sqrt{-t_T}}{\sqrt{4-t_T}-\sqrt{-t_T}}\Bigr)\\
&&\nn\\
B_0(s;0,m_W^2)-B_0(0;m_W^2,m_W^2)&=&2+\Bigl(\frac{1}{s_W}-1\Bigr)~\Bigl\{\log\Bigl(s_W-1\Bigr)-i\pi\Bigr\}\\
B_0(u;0,m_W^2)-B_0(0;m_W^2,m_W^2)&=&2+\Bigl(\frac{1}{u_W}-1\Bigr)~\log\Bigl(1-u_W\Bigr)\\
B_0(t;m_W^2,m_W^2)-B_0(0;m_W^2,m_W^2)&=&2-\sqrt{1-\frac{4}{t_W}}~\log\Bigl(\frac{\sqrt{4-t_W}+\sqrt{-t_W}}{\sqrt{4-t_W}-\sqrt{-t_W}}\Bigr)\\
B_0(m_h^2;m_W^2,m_W^2)-B_0(0;m_W^2,m_W^2)&=&2-2\sqrt{\frac{4}{h_W}-1}~\sin^{-1}\Bigl(\sqrt{\frac{h_W}{4}}\Bigr)\\
&&\nn\\
B_0(s;0,m_Z^2)-B_0(0;0,m_Z^2)&=&1+\Bigl(\frac{1}{s_Z}-1\Bigr)~\Bigl\{\log\Bigl(s_Z-1\Bigr)-i\pi\Bigr\}\\
B_0(u;0,m_Z^2)-B_0(0;0,m_Z^2)&=&1+\Bigl(\frac{1}{u_Z}-1\Bigr)~\log\Bigl(1-u_Z\Bigr)\\
B_0(m_h^2;m_Z^2,m_Z^2)-B_0(0;0,m_Z^2)&=&1-2\sqrt{\frac{4}{h_Z}-1}~\sin^{-1}\Bigl(\sqrt{\frac{h_Z}{4}}\Bigr)~.
\eea

\subsection{Three-point integrals}
We introduce the following parameters: 
\bea
\lambda_1&\equiv&\frac{1}{2}\left(1-\sqrt{1-\frac{4}{t_W}}\right)~,\quad \lambda_2\equiv\frac{1}{2}\left(1+\sqrt{1-\frac{4}{t_W}}\right)\\
x_{W_+}&\equiv&\frac{1}{2}\Bigl(1+i\sqrt{\frac{4}{h_W}-1}\Bigr)~,\quad 
x_{W_-}\equiv\frac{1}{2}\Bigl(1-i\sqrt{\frac{4}{h_W}-1}\Bigr)~\\
x_{Z_+}&\equiv&\frac{1}{2}\Bigl(1+i\sqrt{\frac{4}{h_Z}-1}\Bigr)~,\quad 
x_{Z_-}\equiv\frac{1}{2}\Bigl(1-i\sqrt{\frac{4}{h_Z}-1}\Bigr)~.
\eea
The  three-point integrals given below are all finite except for  the last two.

\bea
C_0(m_h^2,0,t;m_t^2,m_t^2,m_t^2)&=&\frac{1}{t-m_h^2}
\biggl\{\frac{1}{2}\log^2\Bigl(\frac{\sqrt{4-t_T}+\sqrt{-t_T}}{\sqrt{4-t_T}-\sqrt{-t_T}}\Bigr)+2\Bigl[\sin^{-1}\sqrt{\frac{h_T}{4}}\Bigr]^2
\biggr\}\\
&&\nn\\
C_0(0,0,s;m_W^2,m_W^2,0)&=&\frac{1}{s}\Bigl\{\text{Li}_2\Bigl(\frac{1}{s_W}\Bigr) +\frac{1}{2}
\Bigl(\log(s_W)-i \pi\Bigr)^2 +\frac{\pi^2}{6}\Bigr\}\\
C_0(0,0,u;m_W^2,m_W^2,0)&=&-\frac{1}{u}\text{Li}_2\Bigl(u_W\Bigr)\\
C_0(0,0,t;m_W^2,0,m_W^2)&=&\frac{-1}{t}\biggl\{ \text{Li}_2(t_W)+\text{Li}_2\left(\frac{1}{\lambda_1}\right)+\text{Li}_2\left(\frac{1}{\lambda_2}\right)-\text{Li}_2\left(-\frac{\lambda_2}{\lambda_1^2}\right)-\text{Li}_2\left(-\frac{\lambda_1}{\lambda_2^2}\right)\nn\\
&&\hspace{5cm}+4 \log \left(
   -\sqrt{-t_W}~\lambda_1\right) \log (1-\lambda_2
   t_W)
\biggr\}\\
C_0(0,s,m_h^2;m_W^2,0,m_W^2)&=&\frac{-1}{s-m_h^2}\Bigl\{\text{Li}_2\Bigl(\frac{1}{(s_W-h_W)
   x_{W_+}+1}\Bigr)+\text{Li}_2\Bigl(\frac{1}{(s_W-h_W)
   x_{W_-}+1}\Bigr)\nn\\
&&\hspace{-1.5cm}-\text{Li}_2\Bigl(\frac{s_W-h_W+1}{(s_W-h_W)
   x_{W_+}+1}\Bigr)-\text{Li}_2\Bigl(\frac{s_W-h_W+1}{(s_W-h_W)
   x_{W_-}+1}\Bigr)\nn\\
&&\hspace{-1.5cm}-\text{Li}_2(h_W-s_W)+\text{Li}_2\Bigl(\frac{(s_W-1) (s_W-h_W)}{s_W^2-h_W
   s_W+h_W}\Bigr)-\text{Li}_2\Bigl(\frac{h_W-s_W}{s_W^2-h_W
   s_W+h_W}\Bigr)\nn\\
&&\hspace{-1.5cm}+\log (s_W-h_W+1) \log \Bigl(\frac{s_W^2-s_W
   h_W+h_W}{(s_W-h_W)^2}\Bigr)\nn\\
&&\hspace{-1.5cm}+\log (s_W-1) \log \Bigl(\frac{s_W}{s_W^2-s_W
   h_W+h_W}\Bigr)+i \pi  \log \Bigl(\frac{s_W^2-s_W
   h_W+h_W}{s_W}\Bigr)\Bigr\}\\
C_0(0,u,m_h^2;m_W^2,0,m_W^2)&=&\frac{-1}{u-m_h^2}\Bigl\{ \text{Li}_2\Bigl(\frac{1}{1+(u_W-h_W)x_{W_+}}\Bigr)+\text{Li}_2\Bigl(\frac{1}{1+(u_W-h_W)x_{W_-}}\Bigr)\nn\\
&&\hspace{-1.5cm}-\text{Li}_2\Bigl(\frac{u_W-h_W+1}{1+(u_W-h_W)x_{W_+}}\Bigr)-\text{Li}_2\Bigl(\frac{u_W-h_W+1}{1+(u_W-h_W)x_{W_-}}\Bigr)+\text{Li}_2(u_W-h_W+1)\nn\\
&&\hspace{-1.5cm}-\text{Li}_2\Bigl(\frac{u_W}{u_W^2-h_W u_W+h_W}\Bigr)+\text{Li}_2\Bigl(\frac{u_W
   (u_W-h_W+1)}{u_W^2-h_W u_W+h_W}\Bigr)-\frac{\pi ^2}{6}\Bigr\}\\
&&\nn\\
C_0(0,t,m_h^2;m_W^2,m_W^2,m_W^2)&=&C_0(m_h^2,0,t;m_W^2,m_W^2,m_W^2)\nn\\
&=&\frac{1}{t-m_h^2}
\biggl\{\frac{1}{2}\log^2\Bigl(\frac{\sqrt{4-t_W}+\sqrt{-t_W}}{\sqrt{4-t_W}-\sqrt{-t_W}}\Bigr)+2\Bigl[\sin^{-1}\sqrt{\frac{h_W}{4}}\Bigr]^2
\biggr\}\\
&&\nn\\
C_0(0,s,m_h^2;m_Z^2,0,m_Z^2)&=&\frac{-1}{s-m_h^2}\Bigl\{\text{Li}_2\Bigl(\frac{1}{(s_Z -h_Z)
   x_{Z_+}+1}\Bigr)+\text{Li}_2\Bigl(\frac{1}{(s_Z -h_Z)
   x_{Z_-}+1}\Bigr)\nn\\
&&\hspace{-1.5cm}-\text{Li}_2\Bigl(\frac{s_Z-h_Z+1}{(s_Z -h_Z)
   x_{Z_+}+1}\Bigr)-\text{Li}_2\Bigl(\frac{s_Z-h_Z+1}{(s_Z -h_Z)
   x_{Z_-}+1}\Bigr)\nn\\
&&\hspace{-1.5cm}-\text{Li}_2(h_Z-s_Z)+\text{Li}_2\Bigl(\frac{(s_Z-1) (s_Z-h_Z)}{s_Z^2-h_Z
   s_Z+h_Z}\Bigr)-\text{Li}_2\Bigl(\frac{h_Z-s_Z}{s_Z^2-h_Z
   s_Z+h_Z}\Bigr)\nn\\
&&\hspace{-1.5cm}+\log (s_Z-h_Z+1) \log \Bigl(\frac{s_Z^2-s_Z
   h_Z+h_Z}{(s_Z-h_Z)^2}\Bigr)+\log (s_Z-1) \log \Bigl(\frac{s_Z}{s_Z^2-s_Z
   h_Z+h_Z}\Bigr)\nn\\
&&\hspace{-1.5cm}+i \pi  \log \Bigl(\frac{s_Z^2-s_Z
   h_Z+h_Z}{s_Z}\Bigr)\Bigr\}
\\
C_0(0,u,m_h^2;m_Z^2,0,m_Z^2)&=&\frac{-1}{u-m_h^2}\Bigl\{ \text{Li}_2\Bigl(\frac{1}{1+(u_Z-h_Z)x_{Z_+}}\Bigr)+\text{Li}_2\Bigl(\frac{1}{1+(u_Z-h_Z)x_{Z_-}}\Bigr)\nn\\
&&\hspace{-1.5cm}-\text{Li}_2\Bigl(\frac{u_Z-h_Z+1}{1+(u_Z-h_Z)x_{Z_+}}\Bigr)-\text{Li}_2\Bigl(\frac{u_Z-h_Z+1}{1+(u_Z-h_Z)x_{Z_-}}\Bigr)\nn\\
&&\hspace{-1.5cm}+\text{Li}_2(u_Z-h_Z+1)-\text{Li}_2\Bigl(\frac{u_Z}{u_Z^2-h_Z u_Z+h_Z}\Bigr)+\text{Li}_2\Bigl(\frac{u_Z
   (u_Z-h_Z+1)}{u_Z^2-h_Z u_Z+h_Z}\Bigr)-\frac{\pi ^2}{6}\Bigr\}~.
\eea

\bigskip
The following two three-point integrals have collinear divergences which are regularized by 
dimensional regularization. Both integrals are in the form of $C_0(0, 0, p^2; m^2, 0, 0)$, which  corresponds to ``Triangle 3", $I^D_3(0,p_2^2,p_3^2;0,0,m^2)$ with $p_3^2=0$,  of Ref.\cite{EllisZanderighi}. Its expression is  given in Eq.(4.8) of Ref.\cite{EllisZanderighi}.  For   $C_0(0,0,s;m_Z^2,0,0)$,  we need an analytic continuation of the variable $s$ from $s<0$ to $s>m_Z^2$.
\bea
C_0(0,0,s;m_Z^2,0,0)&=&-\Bigl(\frac{4\pi\mu^2}{m_Z^2}\Bigr)^{\epsilon}\frac{1}{s}\Bigl\{~\frac{1}{\epsilon}
\Bigl[\log(s_Z-1)-i\pi\Bigr]-\frac{1}{2}\Bigl[\log(s_Z-1)-i\pi\Bigr]^2\nn\\
&& -\text{Li}_2\Bigl(\frac{s_Z-1}{s_Z}\Bigr)-\frac{1}{2}\log^2\Bigl(\frac{s_Z}{s_Z-1}\Bigr)+\frac{\pi^2}{3}
-i\pi \log\Bigl(\frac{s_Z}{s_Z-1}\Bigr)
\Bigr\}\label{C0Sings}\\
C_0(0,0,u;m_Z^2,0,0)&=&-\Bigl(\frac{4\pi\mu^2}{m_Z^2}\Bigr)^{\epsilon}\frac{1}{u}\Bigl\{~\frac{1}{\epsilon}
\log(1-u_Z)+\text{Li}_2\Bigl(\frac{-u_Z}{1-u_Z}\Bigr)-\frac{1}{2}\log^2(1-u_Z)
\Bigr\}~.\label{C0Singu}
\eea
\bigskip

\subsection{Four-point integrals}

The four-point integrals  $D_0(0,0,0,m_h^2;t,u;m_W^2,0,m_W^2,m_W^2)$ and $D_0(0,0,0,m_h^2;s,t;m_W^2,m_W^2,0,m_W^2)$ appear in Eqs.(\ref{SWk1b}) and (\ref{SWk1bdash}). 
Due to the  relation $$D_0(0,0,0,m_h^2;t,u;m_W^2,0,m_W^2,m_W^2)=D_0(m_h^2,0,0,0;u,t;m_W^2,m_W^2,0,m_W^2),$$ 
these two integrals correspond to the one given in Eq.(37) of Ref.\cite{DNS}. After the analytic continuation procedure  for both dilogarithms  and logarithms~\cite{ EllisZanderighi}, we obtain
\bea
&&  D_0(0,0,0,m_h^2;t,u;m_W^2,0,m_W^2,m_W^2)=D_0(m_h^2,0,0,0;u,t;m_W^2,m_W^2,0,m_W^2)
\nn\\
&&\hspace{1.2cm}=\frac{1}{m_W^4}
\frac{1}{ \sqrt{t_W^2 (u_W-1)^2-2 t_W \left(2 u_W^2-u_W h_W+h_W\right)+h_W^2}}\nn\\
&&\hspace{1.5cm}\times \Biggl\{ 
 \text{Li}_2\left(-\frac{s_1}{r
   x_1}\right)-\text{Li}_2\left(-\frac{s_1}{r
   x_2}\right)+\text{Li}_2\left(-\frac{s_2}{r
   x_1}\right)-\text{Li}_2\left(-\frac{s_2}{r
   x_2}\right)+\text{Li}_2\left(\frac{r
   x_1}{u_W-1}\right)\nn\\
&&\hspace{2cm}+\text{Li}_2\left(\frac{u_W-1}{r
   x_2}\right)-\text{Li}_2\left(\frac{x_1}{u_W-1}\right)
-\text{Li}_2\left(\frac{u_W-1}{x_2}\right)-2
   \text{Li}_2\left(-\frac{1}{x_1}\right)+2
   \text{Li}_2\left(-\frac{1}{x_2}\right)\nn\\
&&\hspace{2cm}+\log (s_2) \log
   \left(-\frac{r x_2+s_1}{r
   x_1+s_1}\right)+\log (s_1) \log \left(-\frac{r
   x_2+s_2}{r x_1+s_2}\right)\nn\\
&&\hspace{2cm}+\log (r)
   \log \left(\frac{x_1^2 (x_2+1)^2
   (u_W-x_1-1)}{ x_2(x_1+1)^2
   (u_W-x_2-1)}\right)\nn\\\nn\\
&&\hspace{2cm}+\log (1-u_W) \log
   \left(\frac{(-u_W+x_1+1) (r
   x_2-u_W+1)}{r (-u_W+x_2+1) (r
   x_1-u_W+1)}\right)+\frac{\log ^2(r)}{2}
\Biggr\}~,
\eea
where
\bea
r&=&1-\frac{1}{2}t_W\left(1+\sqrt{1-\frac{4}{t_W}}\right)~,\qquad s_1=-\frac{x_{W+}}{x_{W-}}~,\qquad s_2= -\frac{x_{W-}}{x_{W+}}\label{r2s1s2}\\
&&\nn\\
x_1&=& \frac{1}{4}\left(\sqrt{1-\frac{4}{t_W}}-1\right)\biggl\{h_W+(1-u_W) t_W\sqrt{1-\frac{4}{t_W}}\nn\\
&&\hspace{3.5cm}+\sqrt{ t_W^2 (1-u_W)^2-2 t_W \left(2 u_W^2-u_W h_W+h_W \right)+h_W^2}\biggr\}\\
x_2&=& \frac{1}{4}\left(\sqrt{1-\frac{4}{t_W}}-1\right)\biggl\{h_W+(1-u_W) t_W\sqrt{1-\frac{4}{t_W}}\nn\\
&&\hspace{3.5cm}-\sqrt{ t_W^2 (1-u_W)^2-2 t_W \left(2 u_W^2-u_W h_W+h_W \right)+h_W^2}\biggr\}~.
\eea
\bigskip
\bea
  &&\hspace{-0.8cm} D_0(0,0,0,m_h^2;s,t;m_W^2,m_W^2,0,m_W^2)=\frac{1}{m_W^4}
 \frac{1}{\sqrt{t_W^2 (s_W-1)^2-2 t_W \left(2 s_W^2-s_W h_W+h_W\right)+h_W^2}}\nn\\
&&\hspace{-0.3cm}\times \biggl\{  -\text{Li}_2\left(\frac{s_W-1}{r
   y_1}\right)-\text{Li}_2\left(\frac{r
   y_2}{s_W-1}\right)+2 \text{Li}_2\left(-\frac{1}{r
  y_1}\right)-2 \text{Li}_2\left(-\frac{1}{r
   y_2}\right)-\text{Li}_2\left(-\frac{s_1}{y_1}\right)+\text
   {Li}_2\left(-\frac{s_1}{y_2}\right)\nn\\
&&-\text{Li}_2\left(-\frac{s_2}{y_1}\right)+\text{Li}_2\left(-\frac{s_2}{y_2}\right
   )+\text{Li}_2\left(\frac{s_W-1}{y_1}\right)+\text{Li}_2\left(
\frac{y_2}{s_W-1}\right)\nn\\
&&+\log (r) \log \left(\frac{y_1
   (r y_2+1)^2 (r y_1-s_W+1)}{y_2^2
   (r y_1+1)^2 (r y_2-s_W+1)}\right)+\log
   (s_1) \log \left(\frac{(s_1+y_2)
   (s_2+y_1)}{(s_1+y_1) (s_2+y_2)}\right)\nn\\
&&+\log
   (s_W-1) \log \left(\frac{r (-s_W+y_1+1) (r
  y_2-s_W+1)}{(-s_W+y_2+1) (r
   y_1-s_W+1)}\right)-\frac{1}{2} \log ^2(r)\nn\\
&&+i \pi  \log
   \left(\frac{
   (-s_W+y_2+1) (r y_1-s_W+1)}{ (-s_W+y_1+1) (r
   y_2-s_W+1)}\right)
\biggr\}~,
\eea
where  $r$, $s_1$ and $s_2$ are given in Eq.(\ref{r2s1s2}) and  
\bea
y_1&=&   \frac{1}{4} \left(\sqrt{1-\frac{4}{t_W}}-1\right)
   \biggl\{-h_W-(s_W-1) t_W
   \sqrt{1-\frac{4}{t_W}}\nn\\
&&\hspace{4cm}+\sqrt{t_W^2(s_W-1)^2-2 t_W \left( 2
   s_W^2-h_W s_W+h_W\right)+h_W^2}\biggr\}\\
y_2&=&  \frac{1}{4} \left(\sqrt{1-\frac{4}{t_W}}-1\right)
   \biggl\{-h_W-(s_W-1) t_W
   \sqrt{1-\frac{4}{t_W}}\nn\\
&&\hspace{4cm}-\sqrt{t_W^2(s_W-1)^2-2 t_W \left( 2
   s_W^2-h_W s_W+h_W\right)+h_W^2}\biggr\}~.
\eea

\vspace{1cm}
The four-point integral $D_0(0,0,0,m_h^2;s,u;m_Z^2,0,0,m_Z^2)$, which appears in Eqs.(\ref{SZk1b}) and (\ref{SZk1bdash}), corresponds to the one given in Eq.(4.37) of Ref.\cite {EllisZanderighi} and also in Eq.(4.13) of Ref.\cite {DennerDittmaier}. It has a collinear singularity. After the analytic continuation procedure  for both dilogarithms  and logarithms~\cite{ EllisZanderighi}, we obtain
\bea
D_0(0,0,0,m_h^2;s,u;m_Z^2,0,0,m_Z^2)&=& D_0(0,0,m_h^2,0;s,u;0,0,m_Z^2,m_Z^2)\nn\\
&&\hspace{-5cm}=\frac{1}{su-m_Z^2(s+u)}\biggl\{\Bigl(\frac{4\pi\mu^2}{m_Z^2}\Bigr)^\epsilon e^{-\epsilon\gamma_E}\times\frac{1}{\epsilon}\biggl[-\Bigl[\log(s_Z-1)-i\pi\Bigr]-\log(1-u_Z)\biggr]\nn\\
&&\hspace{-4cm}  +2
   \text{Li}_2\left(\frac{s_Z-1}{s_Z}\right)-2
   \text{Li}_2\left(-\frac{u_Z}{1-u_Z}\right)-2 \text{Li}_2\left(\frac{1}{(1-s_Z)
   (1-u_Z)}\right)\nn\\
&&\hspace{-4cm}+\text{Li}_2\left(-\frac{x_{Z_+}}{
   x_{Z_-}(1-s_Z)}\right)+\text{Li}_2\left(-\frac{x_{Z_-}}{
   x_{Z_+}(1-s_Z)}\right)+\text{Li}_2\left(1+\frac{x_{Z_+}(1-u_Z)
   }{x_{Z_-}}\right)+\text{Li}_2\left(1+\frac
   {x_{Z_-}(1-u_Z)
   }{x_{Z_+}}\right)\nn\\
&&\hspace{-4cm}+\log
   ^2\left(\frac{s_Z}{s_Z-1}\right)+2 \log \Bigl((s_Z-1) (1-u_Z)\Bigr) \log
   \left(\frac{s_Z+u_Z-u_Z s_Z}{(s_Z-1)
   (1-u_Z)}\right)+2 \log
   (s_Z-1) \log (1-u_Z)\nn\\
&&\hspace{-4cm}+\log ^2(1-u_Z)+\log
   \left(1+\frac{x_{Z_+}}{ x_{Z_-}(1-s_Z)}\right)\left\{\log
   \left(-\frac{x_{Z_+}}{x_{Z_-}}\right)-\log (s_Z-1)\right\} \nn\\
&&\hspace{-4cm}+\log \left(1+\frac{x_{Z_-}}{
   x_{Z_+}(1-s_Z)}\right) \left\{\log
   \left(-\frac{x_{Z_-}}{x_{Z_+}}\right)-\log
   (s_Z-1)\right\}-\frac{2
   \pi ^2}{3}\nn\\
&&\hspace{-4cm}+i \pi  \left[2 \log
   \left(\frac{s_Z}{s_Z+u_Z-u_Z s_Z}\right)+\log
   \left(1+\frac{x_{Z_+}}{ x_{Z_-}(1-s_Z)}\right)+\log
   \left(1+\frac{x_{Z_-}}{ x_{Z_+}(1-s_Z)}\right)\right]\biggr\}~.\label{D0Sing}
\eea

\section{Interference terms \label{Interference}}
We write down the contributions from the interference terms of $\sum_{\rm final~electron~spin}~|A(P_e,P_p)|^2$~,
\bea
&&\hspace{-1cm}\sum_{\rm spin}\Bigl[A_{\gamma\gamma}^*(P_e,P_\gamma)A_{Z\gamma}(P_e,P_\gamma)+A_{\gamma\gamma}(P_e,P_\gamma)A_{Z\gamma}^*(P_e,P_\gamma)\Bigr]\nn\\
&=&\Bigl(\frac{e^2g^2}{16\pi^2}\Bigr)^2\Bigl(\frac{-2}{t-m_Z^2}\Bigr)~F_{\gamma\gamma}~F_{Z\gamma}\biggl\{(f_{Ze}+ P_e)\frac{s^2+u^2}{(s+u)^2}+P_\gamma(f_{Ze}P_e+1)  \Bigl(1-\frac{2u}{s+u}\Bigr)
\biggr\}~,\\
&&\hspace{-1cm}\sum_{\rm spin}\Bigl[A_{\gamma\gamma}^*(P_e,P_\gamma)A_{W\nu_e}(P_e,P_\gamma)+A_{\gamma\gamma}(P_e,P_\gamma)
A_{W\nu_e}^*(P_e,P_\gamma)\Bigr]\nn\\
&=&\Bigl(\frac{e^2g^2}{16\pi^2}\Bigr)^2\frac{-m_W}{2}~F_{\gamma\gamma}~(1-P_e)\frac{1}{s+u}\nn\\
&&\times \biggl\{-s~{\rm Re}\Bigl[S^{{W\nu_e}}_{(k_{1})}(s,t,m_h^2,m_W^2)\Bigr]+u~{\rm Re}\Bigl[S^{{W\nu_e}}_{(k'_{1})}(s,t,m_h^2,m_W^2)\Bigr]
\nn\\
&&\hspace{1cm}+P_\gamma\Bigl( s~{\rm Re}\Bigl[S^{{W\nu_e}}_{(k_{1})}(s,t,m_h^2,m_W^2)\Bigr]+u~{\rm Re}\Bigl[S^{{W\nu_e}}_{(k'_{1})}(s,t,m_h^2,m_W^2)\Bigr]  \Bigr)
\biggr\}~,\\
&&\hspace{-1cm}\sum_{\rm spin}\Bigl[A_{\gamma\gamma}^*(P_e,P_\gamma)A_{Ze}(P_e,P_\gamma)+A_{\gamma\gamma}(P_e,P_\gamma)
A_{Ze}^*(P_e,P_\gamma)\Bigr]\nn\\
&=&\Bigl(\frac{e^2g^2}{16\pi^2}\Bigr)^2\Bigl(-\frac{ m_Z}{8\cos^3\theta_W}\Bigr)~\frac{1}{s+u}~F_{\gamma\gamma}\nn\\
&&\times\biggl\{(f^2_{Ze}+2 P_e f_{Ze}+1)\Bigl\{ s~{\rm Re}\Bigl[S^{Ze}_{(k_{1})}(s,t,m_h^2,m_Z^2)\Bigr]-u~{\rm Re}\Bigl[S^{Ze}_{(k'_{1})}(s,t,m_h^2,m_Z^2)\Bigr] \Bigr\}\nn\\
&&\qquad +P_\gamma(P_e f^2_{Ze}+2  f_{Ze}+P_e)\Bigl\{
s~{\rm Re}\Bigl[S^{Ze}_{(k_{1})}(s,t,m_h^2,m_Z^2)\Bigr]+u~{\rm Re}\Bigl[S^{Ze}_{(k'_{1})}(s,t,m_h^2,m_Z^2)\Bigr]
\Bigr\}
\biggr\}~,\\
&&\hspace{-1cm}\sum_{\rm spin}\Bigl[A_{Z\gamma}^*(P_e,P_\gamma)A_{W\nu_e}(P_e,P_\gamma)+A_{Z\gamma}(P_e,P_\gamma)A_{W\nu_e}^*(P_e,P_\gamma)\Bigr]\nn\\
&=&\Bigl(\frac{eg^3}{16\pi^2}\Bigr)^2\frac{ m_W}{2}\Bigl(\frac{-t}{t-m_Z^2}\Bigr)(1-P_e)(f_{Ze}-1)~F_{Z\gamma}\frac{1}{s+u}\nn\\
&&\times\biggl\{-s~{\rm Re}\Bigl[S^{{W\nu_e}}_{(k_{1})}(s,t,m_h^2,m_W^2)\Bigr]+u~{\rm Re}\Bigl[S^{{W\nu_e}}_{(k'_{1})}(s,t,m_h^2,m_W^2)\Bigr]\nn\\
&&\qquad\qquad +P_\gamma\Bigl\{s~{\rm Re}\Bigl[S^{{W\nu_e}}_{(k_{1})}(s,t,m_h^2,m_W^2)\Bigr]+u~{\rm Re}\Bigl[S^{{W\nu_e}}_{(k'_{1})}(s,t,m_h^2,m_W^2)\Bigr]  \Bigr\}
\biggr\}~,\\
&&\hspace{-1cm}\sum_{\rm spin}\Bigl[A_{Z\gamma}^*(P_e,P_\gamma)A_{Ze}(P_e,P_\gamma)+A_{Z\gamma}(P_e,P_\gamma)A_{Ze}^*(P_e,P_\gamma)\Bigr]
\nn\\
&=&\Bigl(\frac{eg^3}{16\pi^2}\Bigr)^2\Bigl(-\frac{ m_Z}{16\cos^3\theta_W}\Bigr)\Bigl(\frac{1}{t-m_Z^2}\Bigr)~F_{Z\gamma}~\frac{2t}{s+u}\nn\\
&&\times\biggl\{\Bigl( f_{Ze}^3+3f_{Ze}^2 P_e+3f_{Ze}+P_e  \Bigr)
\Bigl\{s~{\rm Re}\Bigl[S^{Ze}_{(k_{1})}(s,t,m_h^2,m_Z^2)\Bigr]-u~{\rm Re}\Bigl[S^{Ze}_{(k'_{1})}(s,t,m_h^2,m_Z^2)\Bigr]\Bigr\}\nn\\
&&\qquad +P_\gamma\Bigl( f_{Ze}^3P_e+3f_{Ze}^2 +3f_{Ze}P_e+1  \Bigr)
\Bigl\{s~{\rm Re}\Bigl[S^{Ze}_{(k_{1})}(s,t,m_h^2,m_Z^2)\Bigr]+u~{\rm Re}\Bigl[S^{Ze}_{(k'_{1})}(s,t,m_h^2,m_Z^2)\Bigr]\Bigr\}
\biggr\}~,\nn\\
&&\\
&&\hspace{-1cm}\sum_{\rm spin}\Bigl[A^{*}_{W\nu_e}(P_e,P_\gamma)A_{Ze}(P_e,P_\gamma)+A_{W\nu_e}(P_e,P_\gamma)
A_{Ze}^*(P_e,P_\gamma)\Bigr]\nn\\
&=&\Bigl(\frac{eg^3}{16\pi^2}\Bigr)^2 \frac{ m_W}{4}\Bigl(-\frac{ m_Z}{16\cos^3\theta_W}\Bigr)
(1-P_e)(f_{Ze}-1)^2(-2t)\nn\\
&&\times \biggl\{{\rm Re}\Bigl\{\Bigl[S^{{W\nu_e}}_{(k_{1})}(s,t,m_h^2,m_W^2)\Bigr]\Bigl[S^{Ze}_{(k_{1})}(s,t,m_h^2,m_Z^2)\Bigr]^*  
+\Bigl[S^{{W\nu_e}}_{(k'_{1})}(s,t,m_h^2,m_W^2)\Bigr]\Bigl[S^{Ze}_{(k'_{1})}(s,t,m_h^2,m_Z^2)\Bigr]^*
\Bigr\}\nn\\
&&\qquad+P_\gamma{\rm Re}\Bigl\{-\Bigl[S^{{W\nu_e}}_{(k_{1})}(s,t,m_h^2,m_W^2)\Bigr]\Bigl[S^{Ze}_{(k_{1})}(s,t,m_h^2,m_Z^2)\Bigr]^*  \nn\\
&&\hspace{6cm}
+\Bigl[S^{{W\nu_e}}_{(k'_{1})}(s,t,m_h^2,m_W^2)\Bigr]\Bigl[S^{Ze}_{(k'_{1})}(s,t,m_h^2,m_Z^2)\Bigr]^*
\Bigr\}
\biggr\}~.
\eea
With the fact that $S^{{W\nu_e}}_{(k'_{1})}$ and $S^{Ze}_{(k'_{1})}$ vanish as $u\rightarrow 0$, it is easy to see that the contributions from the interference terms reduce to zero as $u\rightarrow 0$ 
when $P_e P_\gamma=-1$.
\vspace{2cm}




\end{document}